\def\kbv{$\beta_{BV}$}
\def\kbr{$\beta_{BR}$}
\def\kbi{$\beta_{BI}$}
\def\bbv{$B_{BV}$}
\def\bbr{$B_{BR}$}
\def\bbi{$B_{BI}$}
\def\BVm{$(B_{max}-V_{max})$}
\def\Bm{$B_{max}$}

\def\MBm{$M^B_{max}$}

\def\Mbbv{$M^B_{BV}$}

\def\dm15{$\Delta m_{15}$}
\def\gs{\mathrel{\raise0.27ex\hbox{$>$}\kern-0.70em 
\lower0.71ex\hbox{{$\scriptstyle \sim$}}}}
\def\ls{\mathrel{\raise0.27ex\hbox{$<$}\kern-0.70em 
\lower0.71ex\hbox{{$\scriptstyle \sim$}}}}

\documentclass[12pt,preprint]{aastex}

\usepackage{epsfig}
\usepackage{natbib}
\slugcomment{SN Ia Calibrations}

\shorttitle{Calibrating SN Ia}
\shortauthors{Wang, Goldhaber, Aldering, and Perlmutter}

\begin{document}


\title{
Multi-Color Light Curves of Type Ia Supernovae on the Color-Magnitude 
Diagram: a Novel Step Toward More Precise Distance and Extinction Estimates
      }


\author{Lifan Wang, Gerson Goldhaber\altaffilmark{1}, Greg Aldering, Saul Perlmutter} 
\affil{Lawrence Berkeley National Laboratory 50-232, 1-Cyclotron Rd. CA 94720}
\email{lwang@lbl.gov}


\altaffiltext{1}{Also Physics Department, University of California at 
Berkeley}


\begin{abstract}
We show empirically that fits to the color-magnitude relation of Type Ia
supernovae after optical maximum can provide accurate relative
extragalactic distances.   We report the discovery of an empirical color
relation
for Type Ia light curves:  During much of the first month past maximum,
the magnitudes of Type Ia supernovae defined at a given value of color
index
have a very small magnitude dispersion; moreover, during this period the
relation
between $B$ magnitude and $B-V$ color (or $B-R$ or $B-I$ color) is
strikingly linear, to the accuracy of existing well-measured data.
These linear relations can provide robust distance estimates,
in particular, by using the magnitudes when the supernova reaches
a given color. After correction for light curve strech factor or
decline rate, the dispersion of the magnitudes taken at the
intercept of the linear color-magnitude relation
are found to be around 0$^m$.08 for the sub-sample of supernovae
with \BVm\ $\le\ 0^m.05$, and around 0$^m$.11 for the sub-sample
with \BVm\ $\le\ 0^m.2$. This small dispersion is consistent with
being mostly due to observational errors. The method presented here
and the conventional light curve fitting methods can be combined
to further improve statistical dispersions of distance estimates.
It can be combined with the magnitude at maximum
to deduce dust extinction. The slopes of the color-magnitude relation
may also be used to identify intrinsically different SN Ia systems.
The method provides a tool that is fundamental to
using SN Ia to estimate cosmological parameters such as the Hubble
constant
and the mass and dark energy content of the universe.
\end{abstract}



\section{Introduction}

SNe Ia provide reliable measurements of extragalactic distances. The 
conventional method depends on accurate observations near optical
maxima and detailed light curve shapes to deduce luminosity corrections from
the initial decline rates 
or the time scale stretch factor of the light curves 
\citep{Pskovskii:1977, Phillips:1993, Perlmutter:1997}, or luminosity 
corrections from lightcurve shape fits \citep{Riess:1996}.
Existing observations and theories indicate that SN Ia are indeed 
excellent distance indicators in spite of concerns about several 
systematic effects such as gray dust, evolution and asymmetry.
Relative magnitudes accurate to about $15\%$ have been deduced based on 
supernovae observed in the redshift range from about 3000 km/sec to 
30,000 km/sec \citep{Hamuy:1994, Riess:1996, Phillips:1999}. It 
has also been shown that a two parameter luminosity
correction which employs fits to \dm15\ and \BVm\ can 
further reduce the magnitude dispersion \citep{Tripp:1997, Tripp:1999}.

All of the existing light curve analysis methods aim at 
deriving magnitudes at maximum light and the parameters governing the 
shapes of the observed light curves. The method by \citet{Hamuy:1993, 
Phillips:1993} employs a number of well observed supernova light curves
to model observed light curves to derive simultaneously the magnitude at
maximum and the light curve decline rate $\Delta m_{15}$ (the $B$ magnitude
decline from the peak at 15 days after explosion). Another approach 
called the multi-color light curve shape method (MLCS) 
was introduced by \citet{Riess:1996, Riess:1999} which makes joint
fits to light curves at different colors to obtain simultaneously 
the magnitudes at different colors, the light curve shape corrections to
the magnitudes, and the color-excesses \citep{Riess:1996}. This approach
attempts, within a certain class of models, to statistically determine
the best indicator of luminosity. Alternatively,
\citet{Perlmutter:1997} introduced a time-scale stretch parameter to quantify
the shapes of SN light curves. It was found that stretch describes 
the observed $B$ and $V$ band light curves well over a month around 
maximum \citep{Goldhaber:2001}. Attempts to derive other parameters from
the morphology of Type Ia light curves that
are correlated with the magnitude at optical maximum are reported in
\citet{Hamuy:1996a, Hamuy:1996b}.

While the magnitude at maximum has been used as a standard value, there is 
no physical reason as to why this magnitude is superior to the magnitude at 
other epochs. As shown by the various published analyses, the maximum 
magnitudes can be corrected by a one or two parameter relation to reduce the 
scatter in absolute magnitude to less than 15\%. If SNe Ia are indeed a one 
or two parameter family, it is in principle possible to deduce distance 
measurements from any points on the light curve, provided that required 
parameters for correcting the measured magnitudes can be found, by methods 
such as light curve shapes or spectroscopy (e.g. Nugent et al. 1995; 
Riess et al. 1998).

In this paper, we outline a Color-Magnitude Intercept Calibration (CMAGIC)
method that concentrates on multi-color 
post-maximum light curves of SNe Ia and makes use of their color information. 
We discovered an empirical color relation
for Type Ia light curves:  During much of the first month past maximum,
the magnitudes of Type Ia supernovae defined at a given value of color
index
have a very small magnitude dispersion; moreover, during this period the
relation
between $B$ magnitude and $B-V$ color (or $B-R$ or $B-I$ color) is
strikingly linear, to the accuracy of existing well-measured data.
We will employ the extinctions given by \citet{Phillips:1999} for most 
of this study, but we will also show that our method can provide reliable
independent estimates of extinction. 
We show that accurate calibration can be obtained without data around 
optical maximum. 
This method suggests new observational strategies to obtain accurate distance estimates.
The complete analysis presented in this paper was made for $B$-magnitude 
vs $B-V$ Color. We will also discuss $B-R$ and $B-I$ colors, but will present a more
thorough study in a later paper. We emphasize that
this is a new method for calibrating SN Ia as distance measurements and
more work is in progress. 

\section{The Method}

We analyze the data of the nearby supernova samples published by 
\citet{Hamuy:1996a}, \citet{Riess:1999}, and \citet{Krisciunas:2001}. For
each supernova we study the relationship between $B$ and $B-V$ 
(the ``Color-Magnitude'' or ``CMAG'' plot) as shown in the bottom panels 
of Figures 1-4 for a few example supernovae. The complete set of CMAG plots 
of well-measured supernovae 
are shown in Appendix A. Figures A1, A2, and A3 are for $B$ versus $B-V$,
$B$ versus $B-R$, and $B$ versus $B-I$, respectively. 

The immediate post-maximum CMAG evolution of all these supernovae 
except SN~1992K show a period during which  the observed $B$ 
magnitude and the various colors are linearly related. The examples
shown in Figures 1-4 exhibit some variations in behavior outside this
linear region. In general, it is convenient to write the color-magnitude 
relations as

$$B\ = \ B_{BV}\ + \ \beta_{BV}\ (B-V) \eqno(1a)$$
$$B\ = \ B_{BR}\ + \ \beta_{BR}\ (B-R) \eqno(1b)$$
$$B\ = \ B_{BI}\ + \ \beta_{BI}\ (B-I) \eqno(1c)$$

Here \kbv, \kbr, and \kbi\ give the slopes of the curves, and for linear
$B$ versus $B$-$V$, $B$-$R$, and $B$-$I$ relations, \bbv, \bbr, and \bbi\ are 
constants in time for each fit to an individual supernova. 
The slopes $\beta$ are defined equivalently by

$$ \beta_{BV}\ = \ \dot{B}/(\dot{B}-\dot{V}) \eqno(2a) $$
$$ \beta_{BR}\ = \ \dot{B}/(\dot{B}-\dot{R}) \eqno(2b) $$
$$ \beta_{BI}\ = \ \dot{B}/(\dot{B}-\dot{I}) \eqno(2c) $$

Note that equations (2) are valid not only in the linear regime of the 
CMAG relation but can also be used to describe the non-linear 
parts of the CMAG curves, although this study is dedicated only 
to the linear regime of the CMAG relation. As we will see in the
following sections, the supernovae behave very uniformly in the
linear region, but less so in the before and after regions, 
therefore we will focus only on the linear region in this study.

The linear regime can be explored for its uniformity and applications
to distance determination. Direct extrapolation of the linear fits to 
$(B-V)$ = 0, $(B-R)$ = 0, and $(B-I)$ = 0 would give values 
for intercepts \bbv, \bbr, and \bbi.
However, these values would introduce extrapolation errors correlated with the 
errors in \kbv, \kbr, and \kbi. In practice, such correlated errors can be 
minimized by calculating a magnitude close to the mean of the various 
colors. In this study, we focus only on $B$ and $V$ band data. As we shall see,
the mean color $<B-V>$ is typically around 0$^m.6$ and the $K$-corrected 
$<$\kbv$>$ is typically 1.94. 
For consistency across different studies, it is convenient to define an 
arbitrary reference color that is close to the color average for each pair 
of bands. The intercept of 
$B-V$ = 0$^m$.6 and the line describing the linear region (as given
in equations (1a) and (2a)) defines a magnitude $B_{BV0.6}$ that is 
least correlated to the errors of fitting slope. 
For convenience in comparing to $B_{max}$, we will use 
$B_{BV}\ \equiv \ B_{BV0.6}\ - \ 1.94\times0^m.6 \ = \ \ B_{BV0.6}\ - \ 1^m.164$
throughout this paper as the standard color magnitude intercept.

The fitted values of \bbv\ and \kbv\ are shown in 
Table 1 for all of these supernovae for which more than 3 data points
on the post maximum linear region of the CMAG plot are available. 
A two day gap is introduced around the Transition-Date (cf \S2.1 and 
captions to Figure~1) where the data points are not included in the 
linear fits. Various SNe exhibit different color behavior during 
the epoch before and after the linear region but nearly all show 
very similar behavior during the linear region with small variations 
in the value of the slope. 

\clearpage
\begin{deluxetable}{rrrrrrcrrr}
\tabletypesize{\scriptsize}
\rotate
\tablecolumns{10} 
\tablewidth{0pt}
\tablecaption{Fits to Post-Maximum Light Curves of SN Ia}
\tablehead{
\colhead{SN}           	 & \colhead{$log_(cz)$}\tablenotemark{a}      &
\colhead{B$_{BV}(\sigma)$}    &
\colhead{{B}$^{fix}_{BV}(\sigma)$}\tablenotemark{b}    &
\colhead{$\beta_{BV}(\sigma)$}\tablenotemark{c}		      &
\colhead{B$_{max}$} \tablenotemark{d}	&
\colhead{\BVm}\tablenotemark{d}	&
\colhead{\dm15($\sigma$)}\tablenotemark{d}    &
\colhead{${E_{BV}}(B-V)$($\sigma$)} &
\colhead{Hubble Type\tablenotemark{e}}
}
\startdata 
	(1)   & (2)      & (3)            &(4)              &(5)             & (6)            & (7)      & (8)           &(9)              &(10)                       \\
\hline
     1990O    & 3.957    &16.53    (014 ) & 16.52    (047 ) &  2.07    (02 ) & 16.59    (08 ) &  0.01    & 0.96    (10 ) &  0.15    (03  ) &                       SBa\\
     1990T    & 4.080    &17.22    (010 ) & 17.21    (008 ) &  1.88    (01 ) & 17.27    (20 ) &  0.04    & 1.15    (10 ) &  0.08    (04  ) &                   SA(s)00\\
     1990Y    & 4.070    &16.95    (137 ) & 17.00    (028 ) &  2.05    (12 ) & 17.69    (20 ) &  0.33    & 1.13    (10 ) &  0.13    (06  ) &                   E(M32?)\\
    1990af    & 4.178    &17.74    (037 ) & 17.71    (021 ) &  2.10    (08 ) & 17.91    (03 ) &  0.05    & 1.56    (05 ) &  0.08    (02  ) &                       SB0\\
     1991S    & 4.222    &17.85    (084 ) & 17.87    (031 ) &  1.72    (15 ) & 17.78    (20 ) &  0.04    & 1.04    (10 ) &  0.04    (05  ) &                        Sb\\
     1991U    & 3.968    &16.53    (047 ) & 16.53    (020 ) &  1.90    (05 ) & 16.67    (20 ) &  0.06    & 1.06    (10 ) &  0.10    (04  ) &                   Sbc:pec\\
    1991ag    & 3.617    &14.76    (071 ) & 14.77    (014 ) &  1.96    (07 ) & 14.67    (13 ) &  0.08    & 0.87    (10 ) &  0.09    (04  ) &                SB(s)dmPec\\
     1992J    & 4.137    &17.44    (159 ) & 17.51    (019 ) &  2.08    (14 ) & 17.88    (20 ) &  0.12    & 1.56    (10 ) &  0.10    (04  ) &                         S\\
    1992ae    & 4.351    &18.54    (051 ) & 18.52    (038 ) &  2.08    (08 ) & 18.64    (10 ) &  0.11    & 1.28    (10 ) &  0.09    (03  ) &                       E1?\\
    1992ag    & 3.890    &16.23    (114 ) & 16.21    (068 ) &  1.80    (13 ) & 16.64    (05 ) &  0.13    & 1.19    (10 ) &  0.21    (05  ) &                        S?\\
    1992al    & 3.626    &14.80    (013 ) & 14.72    (020 ) &  2.18    (04 ) & 14.59    (03 ) & -0.05    & 1.11    (05 ) &  0.03    (01  ) &                   SAB(s)c\\
    1992aq    & 4.481    &19.10    (063 ) & 19.16    (033 ) &  1.60    (14 ) & 19.39    (07 ) &  0.10    & 1.46    (10 ) &  0.12    (04  ) &                       Sa?\\
    1992au    & 4.261    &18.13    (255 ) & 18.09    (046 ) &  1.71    (32 ) & 18.17    (20 ) &  0.05    & 1.49    (10 ) &  0.02    (06  ) &                        E1\\
    1992bc    & 3.773    &15.61    (038 ) & 15.62    (017 ) &  1.93    (05 ) & 15.15    (03 ) & -0.08    & 0.87    (05 ) &  0.02    (01  ) &                       Sab\\
    1992bg    & 4.029    &17.19    (031 ) & 17.18    (026 ) &  2.11    (05 ) & 17.39    (05 ) & -0.04    & 1.15    (10 ) &  0.20    (03  ) &                        Sa\\
    1992bh    & 4.131    &17.47    (037 ) & 17.49    (030 ) &  2.05    (04 ) & 17.68    (05 ) &  0.08    & 1.05    (10 ) &  0.18    (03  ) &                       Sbc\\
    1992bk    & 4.240    &18.06    (030 ) & 18.07    (044 ) &  1.64    (05 ) & 18.07    (08 ) &  0.00    & 1.57    (10 ) &  0.01    (04  ) &                    S0-pec\\
    1992bl    & 4.110    &17.36    (055 ) & 17.34    (036 ) &  2.05    (09 ) & 17.34    (05 ) &  0.00    & 1.51    (10 ) &  0.01    (03  ) &              (R1)SB(s)0/a\\
    1992bo    & 3.735    &15.78    (036 ) & 15.78    (009 ) &  1.99    (06 ) & 15.85    (03 ) &  0.01    & 1.69    (05 ) &  0.03    (01  ) &                SB(s)00pec\\
    1992bp    & 4.374    &18.41    (057 ) & 18.46    (045 ) &  1.72    (11 ) & 18.53    (03 ) & -0.05    & 1.32    (10 ) &  0.11    (03  ) &                     E2/S0\\
    1992br    & 4.420    &19.14    (055 ) & 19.17    (076 ) &  1.64    (08 ) & 19.34    (16 ) &  0.04    & 1.69    (10 ) &  0.05    (05  ) &                        E0\\
    1992bs    & 4.279    &18.33    (047 ) & 18.32    (030 ) &  2.02    (08 ) & 18.33    (06 ) &  0.07    & 1.13    (10 ) &  0.08    (03  ) &                   Sc(s)II\\
     1993B    & 4.326    &18.36    (055 ) & 18.43    (033 ) &  1.61    (12 ) & 18.81    (09 ) &  0.12    & 1.04    (10 ) &  0.24    (05  ) &                       SBb\\
     1993H    & 3.862    &16.30    (051 ) & 16.31    (021 ) &  2.00    (06 ) & 16.99    (05 ) &  0.23    & 1.69    (10 ) &  0.24    (04  ) &                   SB(r)ab\\
     1993O    & 4.192    &17.76    (055 ) & 17.76    (023 ) &  2.08    (07 ) & 17.79    (03 ) & -0.09    & 1.22    (05 ) &  0.09    (02  ) &                    E5/S01\\
    1993ac    & 4.169    &18.14    (029 ) & 18.17    (026 ) &  2.18    (03 ) & 18.45    (15 ) &  0.20    & 1.19    (10 ) &  0.20    (04  ) &                         E\\
    1993ae    & 3.734    &15.50    (115 ) & 15.54    (037 ) &  2.10    (13 ) & 15.40    (20 ) & -0.07    & 1.43    (10 ) &  0.04    (05  ) &                         E\\
    1993ag    & 4.176    &17.82    (046 ) & 17.82    (025 ) &  2.01    (06 ) & 18.30    (07 ) &  0.03    & 1.32    (10 ) &  0.24    (04  ) &                    E3/S01\\
    1993ah    & 3.947    &16.52    (072 ) & 16.50    (063 ) &  1.65    (09 ) & 16.32    (20 ) & -0.04    & 1.30    (10 ) &  0.02    (05  ) &                         E\\
     1994M    & 3.860    &16.32    (074 ) & 16.27    (039 ) &  2.10    (22 ) & 16.35    (05 ) &  0.05    & 1.44    (10 ) &  0.04    (03  ) &                         E\\
     1994Q    & 3.939    &16.39    (056 ) & 16.39    (019 ) &  1.95    (10 ) & 16.40    (19 ) & -0.04    & 1.03    (10 ) &  0.06    (04  ) &                        S0\\
     1994S    & 3.686    &15.06    (028 ) & 15.06    (013 ) &  1.97    (03 ) & 14.79    (07 ) & -0.04    & 1.10    (10 ) &  0.02    (02  ) &                       Sbc\\
    1994ae    & 3.208    &13.31    (040 ) & 13.31    (012 ) &  1.96    (06 ) & 13.21    (07 ) &  0.11    & 0.86    (05 ) &  0.10    (03  ) &                        Sc\\
     1995D    & 3.293    &13.59    (023 ) & 13.53    (024 ) &  2.33    (05 ) & 13.44    (07 ) &  0.02    & 0.99    (05 ) &  0.08    (02  ) &                        S0\\
     1995E    & 3.540    &15.33    (081 ) & 15.33    (020 ) &  1.95    (07 ) & 16.82    (07 ) &  0.73    & 1.06    (05 ) &  0.52    (04  ) &                        Sb\\
    1995ac    & 4.166    &17.21    (027 ) & 17.19    (016 ) &  2.07    (05 ) & 17.19    (07 ) &  0.08    & 0.91    (05 ) &  0.13    (03  ) &                         S\\
    1995ak    & 3.824    &16.05    (068 ) & 16.04    (025 ) &  1.90    (07 ) & 16.28    (10 ) &  0.10    & 1.26    (10 ) &  0.12    (04  ) &                       Sbc\\
    1995al    & 3.188    &13.25    (124 ) & 13.23    (053 ) &  1.78    (15 ) & 13.36    (07 ) &  0.12    & 0.83    (05 ) &  0.12    (05  ) &                 SA(rs)bc:\\
    1995bd    & 3.681    &15.99    (130 ) & 16.10    (027 ) &  2.14    (10 ) & 17.27    (10 ) &  0.79    & 0.84    (05 ) &  0.62    (07  ) &                         S\\
     1996C    & 3.955    &16.66    (019 ) & 16.65    (018 ) &  1.83    (02 ) & 16.59    (10 ) &  0.07    & 0.97    (10 ) &  0.07    (04  ) &                        Sa\\
     1996X    & 3.308    &13.17    (026 ) & 13.19    (033 ) &  2.15    (03 ) & 13.26    (07 ) &  0.05    & 1.25    (05 ) &  0.11    (02  ) &                        E0\\
     1996Z    & 3.357    &13.84    (053 ) & 13.84    (031 ) &  2.46    (08 ) & 14.61    (10 ) &  0.36    & 1.22    (10 ) &  0.28    (04  ) &                        Sb\\
    1996bl    & 4.019    &16.89    (033 ) & 16.88    (018 ) &  1.87    (03 ) & 17.08    (07 ) &  0.15    & 1.17    (10 ) &  0.17    (03  ) &                       SBc\\
    1996bo    & 3.714    &15.32    (068 ) & 15.33    (021 ) &  1.99    (07 ) & 16.15    (10 ) &  0.41    & 1.25    (05 ) &  0.31    (04  ) &                        Sc\\
    1996bv    & 3.700    &15.26    (053 ) & 15.26    (027 ) &  2.01    (08 ) & 15.77    (13 ) &  0.25    & 0.93    (10 ) &  0.26    (05  ) &                      Scd:\\
    1999dk    & 3.618    &14.79    (092 ) & 14.78    (034 ) &  1.92    (14 ) & 15.06    (10 ) &  0.15    & 1.00    (10 ) &  0.16    (05  ) &                       ???\\
    2000bk    & 3.902    &16.55    (031 ) & 16.56    (011 ) &  1.99    (03 ) & 16.98    (20 ) &  0.17    & 1.63    (10 ) &  0.08    (04  ) &                       ???\\
    2000ce    & 3.695    &16.12    (141 ) & 16.16    (040 ) &  2.01    (11 ) & 17.24    (20 ) &  0.61    & 0.99    (10 ) &  0.22    (06  ) &                       ???\\

\enddata

\tablecomments{The columns are: (1) name of the supernova; (2) redshift; (3)
the Color-Magnitude with errors in units of 0$^m$.001; (4) the Color-Magnitude
for fixed slope fit with errors in units of 0$^m$.001; (5) the fitted slope
with error in units of 0.01; (6) the magnitude at maximum as reported in
\citet{Hamuy:1996a, Riess:1996, Krisciunas:2001}, with errors in 0$^m$.01; (7)
the color at maximum; (8) \dm15\ as reported in 
\citet{Hamuy:1996a, Riess:1996, Krisciunas:2001}; (9) the host galaxy 
extinction derived from CMAGIC after applying the Bayesian prior as
described in \citet{Phillips:1999}; (10) host galaxy type}

\tablenotetext{a}{Heliocentric redshift}
\tablenotetext{b}{Cmag for fixed slope \kbv\ $=\ 1.94\ \pm\ 0.16$.}
\tablenotetext{c}{K-corrected slopes}
\tablenotetext{d}{Numbers from Hamuy et al. 1993, and Riess et al. 1999}
\tablenotetext{e}{As given in Ivanov, Hamuy, \& Pinto 2000}

\end{deluxetable}

Both colors and magnitudes are subject to observational errors. The errors on
$B$-$V$ colors are not available for most of the supernovae; when unavailable
they are constructed
from the published errors of $B$ and $V$ by assuming uncorrelated errors
and are calculated by the square root of the sum of 
the square of the errors on $B$ and $V$ for each SN. This is normally an overestimate 
of the errors in most cases, and the errors thus constructed are correlated
to those at individual colors. These errors
were taken into account in the linear fits and indeed, in most cases the 
resulting $\chi^2$ per degree of freedom values were less than one. 
Since it is hard to track the exact nature of the
correlations of the errors we therefore scaled the color and magnitude 
errors to make $\chi^2_\nu \ = \ 1$. We note that this does not change 
the fitting parameters significantly, but only the final errors associated 
with them. Alternatively, one can also fit linear relations between
$B$ and $V$ directly and then convert the fitted parameters to $B_{BV}$. 
We found no significant differences between the two approaches and will
use the $B$ vs $B-V$ fits for clarity. 

\begin{figure}
\figurenum{1}
\epsscale{0.6}

\plotone{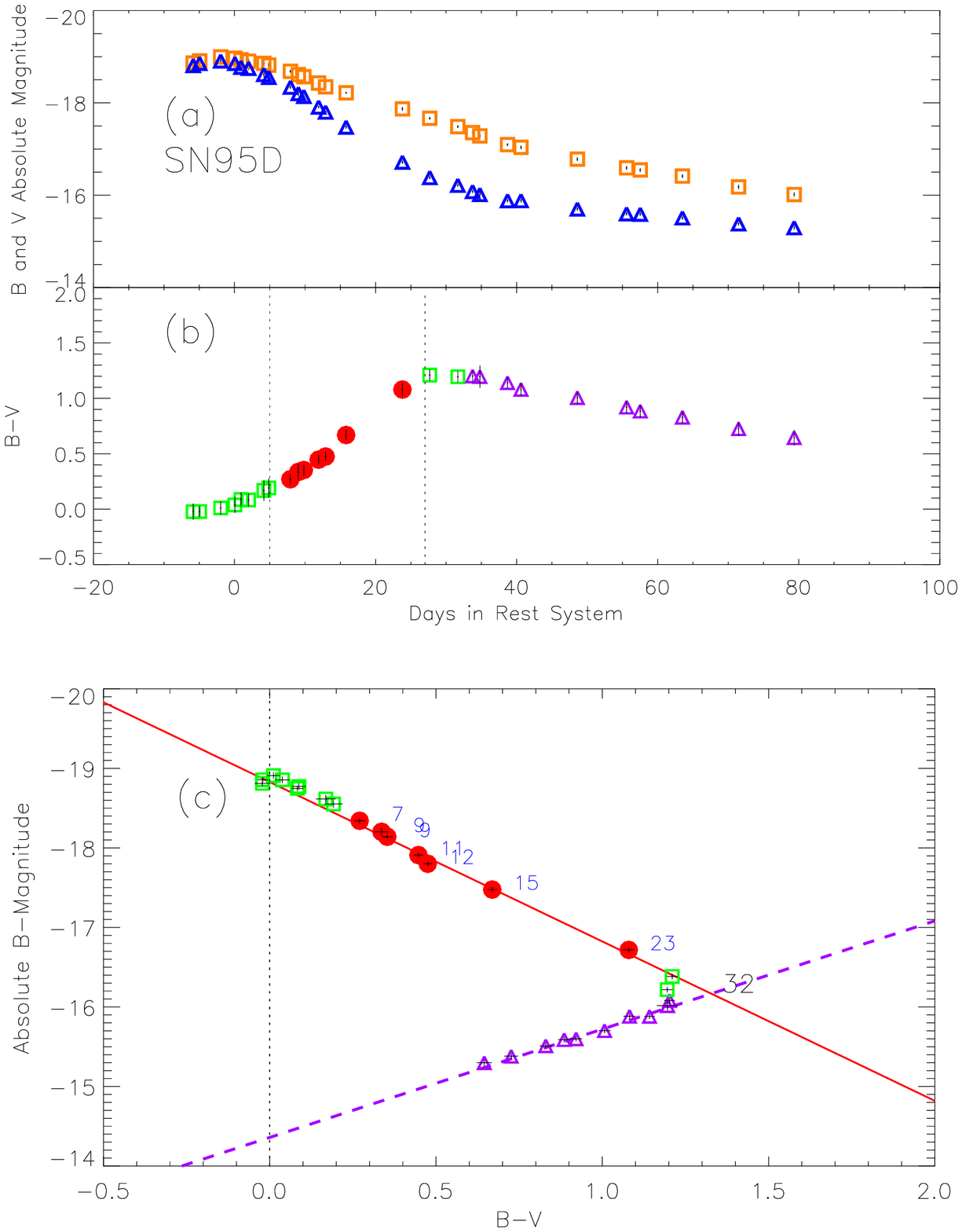} 
\caption{
 Example of a SNe that 
 is typical of the majority of SNe in our sample. 
(a) The lightcurve in the B-band (blue triangles) and V-band 
(orange squares).(b) Color
 vs. Rest-frame day plot. The two vertical lines bracket the linear epoch.
The green squares and purple triangles show data outside
the linear region. The linear region is shown in red
circles.
(c) The Color-Magnitude plot. The epoch before and soon after 
the linear region is shown in green open squares. The epoch of the 
linear region is shown in red filled circles. The epoch after the 
linear region which corresponds to 
the nebular phase is shown in violet open triangles. For the linear 
region the rest frame days after $B$ maximum are given by the blue 
numbers. The red line is the CMAGIC fit to the linear region
 with slope \kbv. The violet dashed line is the fit to the nebular 
phase region. The intersection of these two lines is marked with the 
day (in black) of the abrupt change in color evolution. 
}

\end{figure}
\begin{figure}
\figurenum{2}
\epsscale{0.6}
\plotone{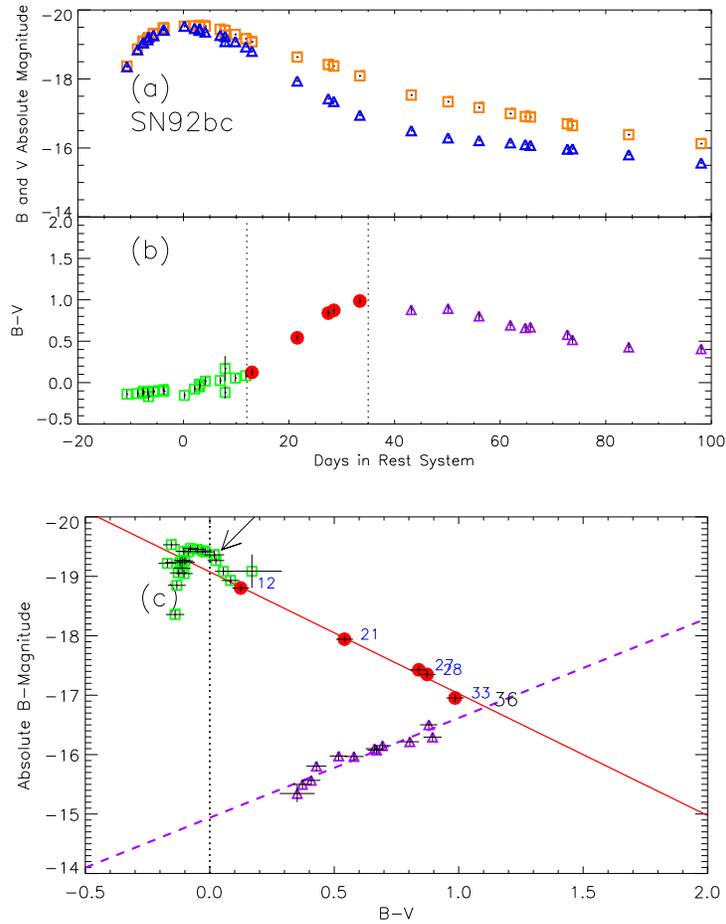} 
\caption{Example of a SNe that exhibits the ``bump'' feature (cf. \S2.1). 
The bump is marked by the arrow in the figure.
This SN which is very bright and has high stretch, low \dm15\ and low extinction.
  See caption to Figure 1.}
\end{figure}

\begin{figure}
\figurenum{3}
\epsscale{0.6}
\plotone{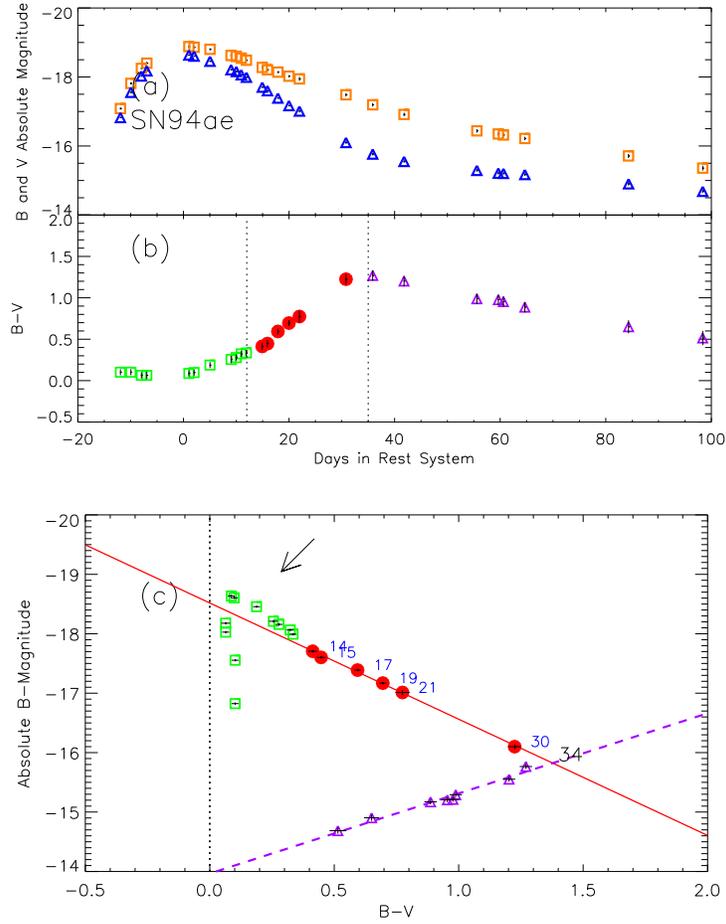} 
\caption {Example of another  SNe that  exhibits the ``bump'' feature. 
The bump is marked by the arrow in the figure.
This SN has also a low \dm15, with large dust extinction however.
See caption to Figure 1.}
\end{figure}

\begin{figure}
\figurenum{4}
\epsscale{0.6}
\plotone{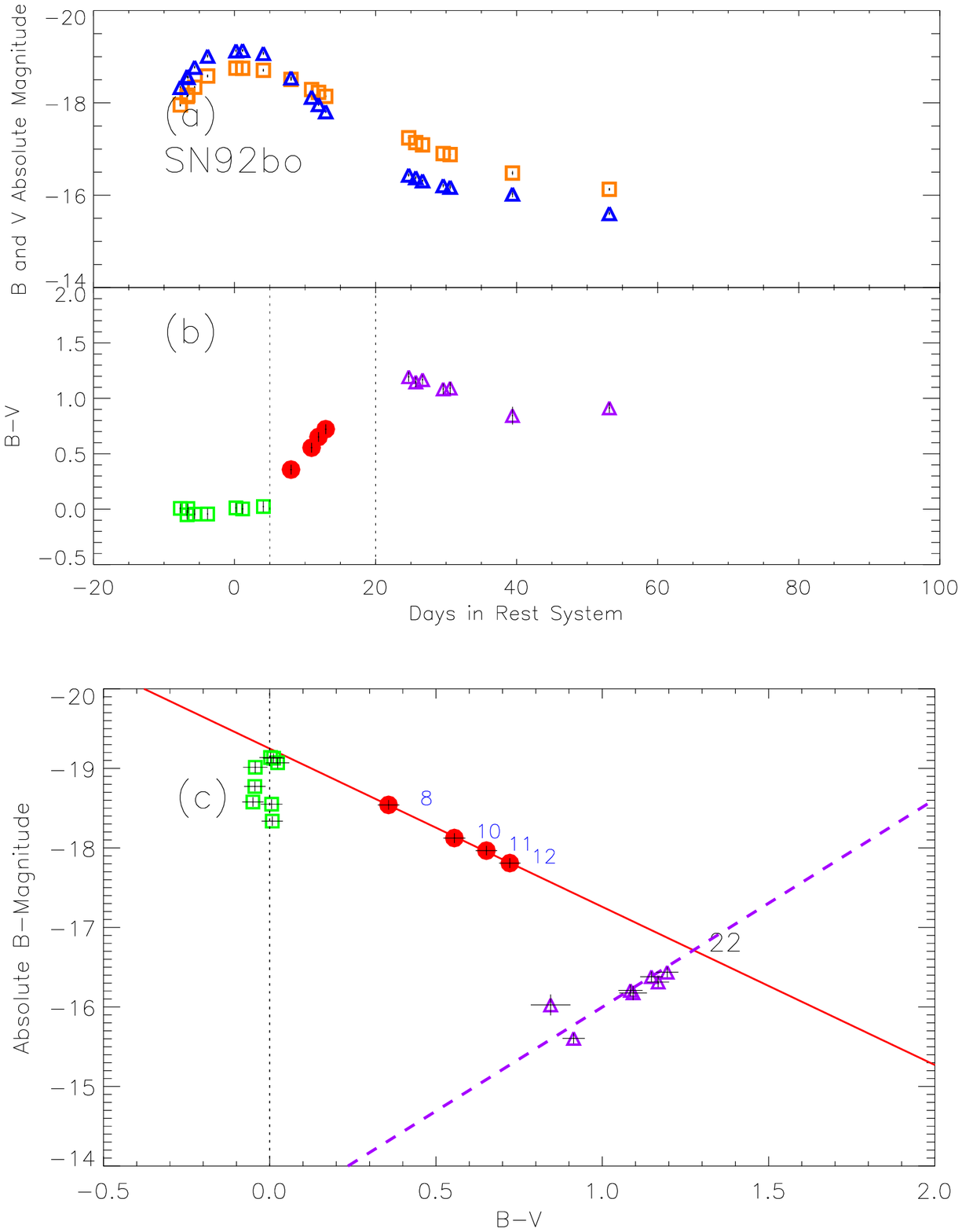}
\caption{A second example of a SN that does not exhibit the ``bump''
 feature. This SN is very dim with large \dm15.  
The Transition-Date is earlier than for supernovae with smaller \dm15.
See caption to Figure 1.}
\end{figure}
    
SN 1992K is the {\it only} supernova in this sample which does not show a 
linear relation with \kbv\ around 2. In fact, the $B-V$ and $B-I$ colors never 
exhibited a redward evolution during the course of the observations. It is an
intrinsically dim SN Ia with a
very fast light curve ($\Delta m_{15}  \ = \ 1^m.93 $) and spectral 
evolution that is very similar to that of SN 1991bg \citep{Hamuy:1994}. This
supernova was discovered after optical maximum. It is likely that the
linear regime was either not observed, or does not exist for this particular
peculiar supernova. For this reason, SN 1992K is excluded from 
further consideration. Such peculiar
supernovae are easy to identify and can be excluded by the shape of the 
color-magnitude curve. 

\subsection{The Shapes of the Color-Magnitude Diagrams of Type Ia Supernovae}

In Figures 1 to 4 we show a number of well measured SNe as examples to 
illustrate the CMAGIC  method for the $B$-band with $B$-$V$ color. 
All data points shown in Figures 1 to 4 have been K-corrected to 
the rest system. In these 
Figures (a) we first plot the $B$ and $V$ lightcurves in absolute magnitudes 
(assuming $H_0$ = 65 km/sec/kpc). The color, for each set of data points, 
is then the difference between the two curves. In (b) we show the color vs 
the epoch in the SN rest frame. In (c) we show the absolute $B$-magnitude vs 
the color $B-V$. The linear regime is identified on the CMAG
plot (c) and is marked by red points. As we will discuss, the epoch over 
which the linear region occurs varies in its starting and ending day, and 
these dates are correlated with the stretch factor or the rate of 
magnitude decline $\Delta m_{15}$. We have shown the early data before and around maximum 
light in green, the linear region in red and the tail or nebular phase region 
in violet. We have indicated the day for each data point for the
linear region in blue. We have 
fitted the linear region (red data points) to obtain the slope \kbv. We have 
also fitted the nebular phase region shown in violet in Figures 1 to 4, 
which is also linear in this representation. 
The intersection of these two lines is marked with the day (in black) at 
which the color evolution changes abruptly from $B-V$ increasing linearly 
towards the red to decreasing linearly. For comparison,  
\citet{Lira:1995, Riess:1996, Phillips:1999} noticed a different 
linear relation between $B$ magnitude versus time during the nebular phase. 
Further analysis of supernovae well observed both around 
optical maxima and up to the nebular phase shows that \dm15\ are strongly 
correlated with the date the supernovae reaches maximum (reddest) $B-V$ 
values. This correlation can be seen in plots given in \citet{Phillips:1999}
and \citet{Garnavich:2001}. Details of this will be presented in another 
separate paper.

 As shown in Figures 1-4, and A1, the evolution of a typical Type Ia supernova 
normally starts at $B-V$ around 0 or bluer before optical maximum, and 
it evolves rapidly to the red soon after optical maximum. 
In the tail region or during the ``supernebular'' phase the CMAG 
plot takes a sharp turn and then evolves rapidly towards 
the blue. This sharp turn defines another characteristic time -- the 
Transition-Date at which the supernova reaches the nebular phase.
This date corresponds approximately to the "inflection point" on the
light curves of SN Ia discussed in \citet{Pskovskii:1977}, 
\citet{Pskovskii:1984}, and \citet{Pavlyuk:1998}, or the "intersection
point" in \citet{Hamuy:1996b}.
The morphology of the evolutionary track is determined by 
the derivatives given in equations (2). The shapes of the CMAG relations 
are therefore insensitive to extinction, redshift, and the stretch values 
of the light curves if $s_B\ \propto\ s_V$, where $s_B$ and $s_V$ are 
the stretch parameters in the $B$ and $V$ bands, respectively. 
Extinction only displaces the entire CMAG curve but does not change 
its shape. Redshift stretches the time coordinates, by $1+z$, equally in all 
colors and does not change the slopes given in equation (2). 
The shapes of the color-magnitude plots are dependent, but only
indirectly, on redshift and extinction via K-correction. These properties
as well as the simple linearity with very small dispersions make CMAG a 
powerful tool for studying intrinsic properties of supernovae.

We derive from Figures 1-4, and A1-3 the CMAGIC linear relation of Type Ia supernovae:
{\it linear relations 
between $B$ magnitude and $B-V$, $B-R$, and $B-I$ colors, universal to
SN~Ia during much of the first month past maximum exist for Type Ia 
supernovae. }

The variations in CMAG curves outside the linear region indicate 
that Type Ia light curves are {\it not} a one parameter family.
In a one parameter family, a supernova is characterized by a single variable
such as the stretch factor, or \dm15. In previous studies, it was found that 
the stretch factor of different colors are correlated. When the extended 
\citet{Leib:1988} templates are used, a relation $s_V\ = \ s_B$ 
\citep{Perlmutter:1997, Perlmutter:1999, Goldhaber:2001} was 
found to approximately describe the Type 
Ia light curves for matching $B$ and $V$ templates. Real scatter on 
the $s_B$ versus $s_V$ diagram was indeed 
observed but was difficult to distinguish from observational errors. 
The CMAG diagram provides a new way to investigate this 
issue. If a fixed linear relation exists for stretch at different 
colors and the light curve at each color is universal to all the 
supernovae with $s_B$ the
only free parameter, to first order, we would expect the CMAG curves and \kbv\ 
as defined in equations (2) to be universal as well. This would imply that 
regardless of the light curve shapes, all the Type Ia supernovae should 
show identical morphology on the CMAG diagram.

The CMAG diagrams shown in Figures 1 to 4 and Figures A1, A2, 
and A3 appear not to agree with the single parameter model outside the
linear region. 
On these $B$ versus $(B-V)$ graphs, the $B$ magnitudes increase rapidly while 
the colors show relatively small evolution until the supernova reaches $B$ 
maximum. After that, the supernovae can be divided into at least two groups. 
The first group shows a ``bump'' of excess luminosity (typically before 
day 12) above the linear fits at around the maximum, and the other does not.

Examples of the SN Ia showing a bump are: SN 1992bc, SN 1994ae, SN 1995al, 
SN 1995bd, SN 1996C, and SN 1996ab. Members of this group are among 
the brightest Type Ia supernovae. Spectroscopic data identified 
1995bd as a SN 1991T-like event \citep{Garnavich:1995}. The 
$\Delta m_{15}$ values for these supernovae are around 0.87 -- the lowest 
end of the templates used to determine these parameters \citep{Phillips:1999, 
Hamuy:1996b, Riess:1996}. For this group of supernovae, 
the epochs at which the linear regimes apply are typically from 12 to 
35, 11 to 33, and 11 to 32 days after optical maximum in the $B$ versus 
$B-V$, $B$ versus $B-R$, and $B$ versus $B-I$ diagrams, respectively. 
SN~1991ag and SN 1992P are the only two supernovae, in the 
sample studied, with \dm15\  about 0.87 for which the 
$B$ versus $(B-V)$ graphs do not show obvious bumps. This seems to be 
due mostly to the incomplete observation time coverages. For SN 1991ag, 
a hint of a bump can in fact be 
seen when a linear CMAG fit is applied to data between day 12 to 35. 
No color information is available between day 12 to 35 for SN 1992P. 
This group of supernovae includes also SN 1991T, SN 1997br \citep{Li:1999}, 
and SN 2000cx \citep{Li:2001}, not in our sample but well observed. 
However, it should be noted that the 
spectroscopically SN 1991T-like event - SN 1995ac \citep{Filippenko:1995} 
does not show the bump that defines this group. 

Most of the supernovae fall into the second group for which the 
CMAG diagrams are linear and without a bump. Typical epochs of the 
linear region
ranges from day 5 to 27, 11 to 31, and 11 to 32 for $B$ versus $(B-V)$, 
$B$  versus $ (B-R)$, and $B$ versus $ (B-I)$, respectively. But some 
supernovae such 
as SN 1992bk, SN 1992bo, and SN 1993ae show a considerably shorter duration 
of the linear region which lasts only about 16 days from day 5 to 20 for 
$B$ versus $B-V$. 
These are typically intrinsically dimmer supernovae with values of \dm15\ above 1.4. 
We note that the peculiar SN~1991bg and SN 1999by \citep{Garnavich:2001} show linear regions
on CMAG plot similar to the one shown in Figure 4. 
Other supernovae with comparable \dm15\, for which however not enough late 
epoch data are available to constrain their evolution around day 27, are 
SN 1992aq, SN 1992au, SN 1992br, SN 1993H, SN 1994M, and SN 1996bk. Their 
behavior on the CMAG diagram are similar. 

In Figure 5 we show the composite residuals of the linear fits 
rebinned to $B-V$ intervals of 0.1. The value at each interval is given 
by the weighted 
average of all the residuals of the linear fits within that bin. Errors 
on both $B$ and $B-V$ were taken into account in these calculations. 
No systematic deviations from a straight line are observed down to the
0$^m.01$ level to which this data is sensitive. For most of the 
very well observed supernovae, the fits usually give $\chi^2$ per 
degree of freedom well below 1, and typical scatter around the
fits is less than 0$^m$.05. 
This demonstrates that on average linear fits indeed provide an accurate 
description of the observed data. At least for the
currently available data, higher order fits, or more complicated color-magnitude
templates are not necessary. 

\begin{figure}
\figurenum{5}
\epsscale{1.0}
\plotone{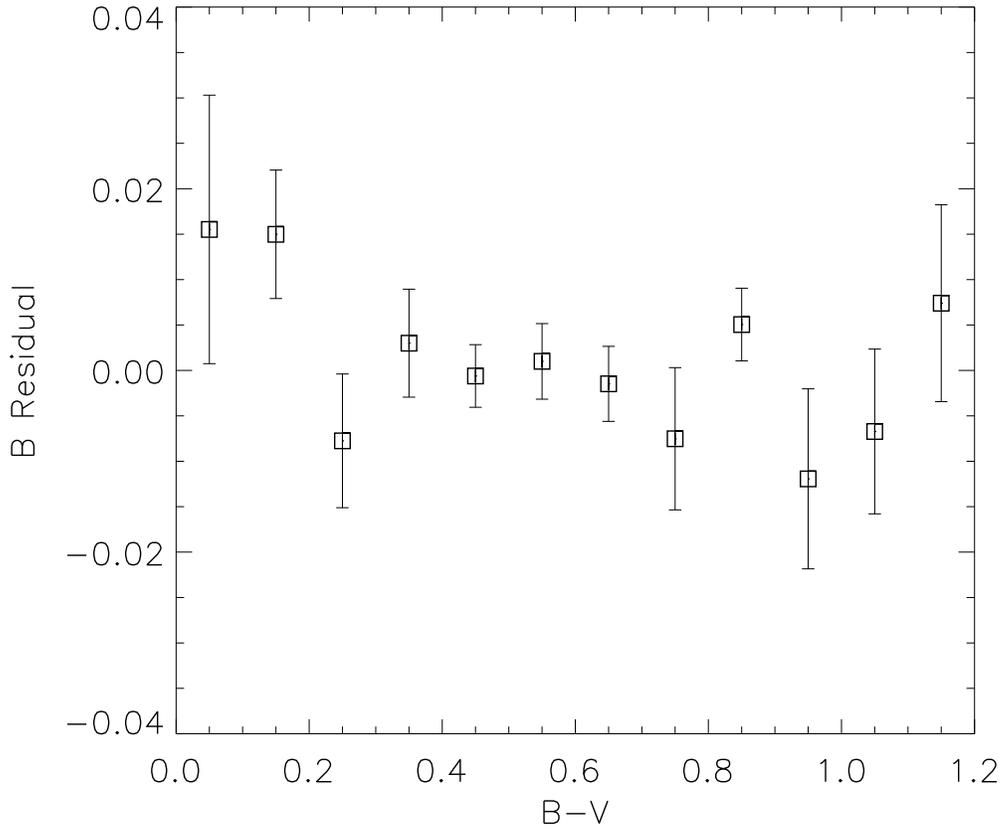}
\caption{Mean composite residuals of the linear fits to the $B$ versus 
$B-V$ curves, binned to 
$B-V$ interval of 0.1. The small residual values indicate that the data
can be very well described by a straight line.
}
\end{figure}

\begin{figure}[t]
\figurenum{6}
\epsscale{1.0}
\plottwo{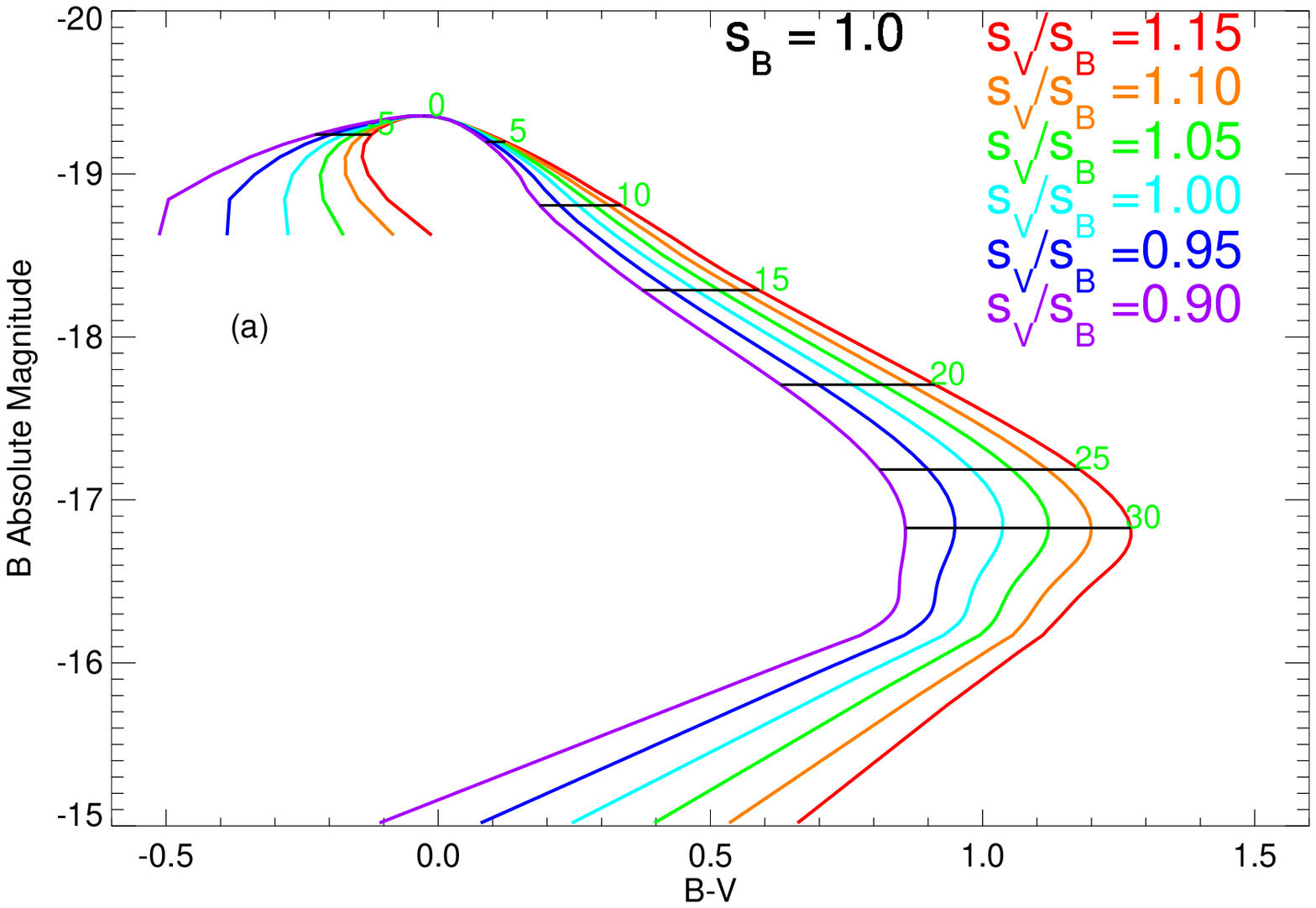}{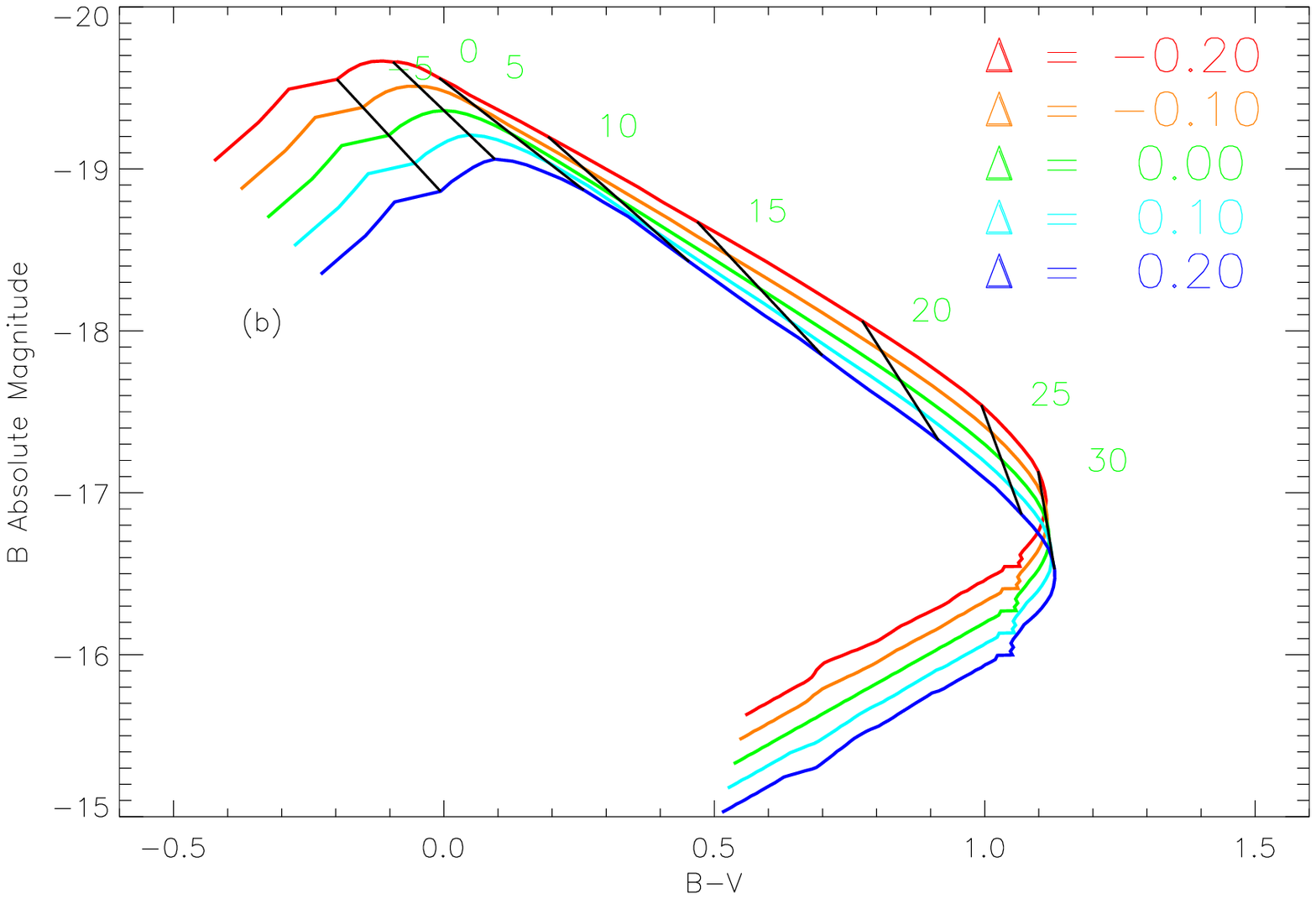}
\caption{(a) Absolute Magnitude versus Color plot from $B$ and $V$ standard templates 
each stretched by $s_B$ and $s_V$ respectively. The small numbers correspond 
to the epoch of the light curve in days from $B$ maximum. In this 
illustrative example $s_B$ = 1.0 is kept fixed while $s_V$ varies 
from 0.90 to 1.15. Reading from left to right at day 15, 
the curves correpond to different stretch ratios 
from $s_V/s_B$ = 0.90 to 1.15. Note that
the linear relations are qualitatively reproduced, and that 
both the light curves without a bump and with a bump 
are qualitatively reproduced from differences of stretched 
standard templates \citep{Leib:1988,Perlmutter:1997,  Perlmutter:1999, 
Goldhaber:2001}. 
(b) The color curves obtained from the templates
of \citet{Riess:1996} with varying correction coefficients
$\Delta$. Reading from bottom to top at day 15, 
the different curves correpond to $\Delta$ values from 0$^m$.2 to -0$^m$.2 as 
shown in the figure. 
The linear relations are well reproduced, but the templates
fail to reproduce any supernovae with a bump.}
\end{figure} 

A possible elaboration of the stretch fitting model to account for the
the behavior on the CMAG Diagram outside the linear region 
can be constructed by assuming a series of different ratios 
of the $B$ and $V$ stretches  $s_B/s_V$. 
This concept is shown in Figure 6~(a), where curves are given for a number of
stretch ratios $s_B/s_V\ =\ 0.90\ -\ 1.15$. The template light curves are the 
$B$-template 'Parab20' discussed in \citet{Goldhaber:2001}, the 
$V$-templates were treated in a similar fashion. Indeed, the morphology 
classes of the CMAG Diagram are qualitatively reproduced. Those with
a bump typically have $B$ stretch larger than $V$ stretch values, 
whereas those without a bump have $B$ stretch values smaller or comparable 
to the $V$ stretch values. It seems that stretch values measured with 
lightcurves of different colors have the potential of measuring more subtle 
differences of the light curve properties. 
However, it should be noted that Figure 6~(a) does not provide quantitative 
fits to the observed morphology shown in Figures 1 to 4, and A1. 
The epochs of the linear regimes are inconsistent with the observed 
values. 

For comparison, we plot in Figure 6~(b) the template spectra constructed by
\citet{Riess:1996} with varying correction coefficients, $\Delta$. The linear 
relation is successfully reproduced, but the templates fail to reproduce
the bumps normally associated with the brightest supernovae. For 
these supernovae, 
this may lead to systematic errors in the color estimates near maximum 
which in turn lead to errors on extinction estimates.

$K$-correction \citep{Oke:1968, Hamuy:1993, Kim:1996, 
Nugent:2002} affects the color magnitudes, and were applied to the 
observed magnitudes before the linear color-magnitude fits. 
We do not expect significant K-correction errors as the redshifts of the 
SN Ia are low ($z$ $\lesssim$ 0.1). 
The linearity of $B$ versus $B-V$, $B-R$, and $B-I$ relations implies 
that the $K$-correction would likely conserve the linearity of the linear 
section of the CMAG relation. At these redshifts, 
an approximate linear relation between the $K$-corrections and the 
observed colors can be constructed. Therefore for the CMAGIC method, 
$K$-corrections may be applied either before or after the linear fits 
to the observed data points. In this study, we have made the
$K$-corrections after the linear CMAG fits of the observed data points.

\subsection{The Characteristics of the Slope $\beta$}

It is found from the relatively well calibrated sample that the mean 
slopes without $K$-corrections are: $<$\kbv $>\ =\  2.07\pm0.18$, 
$<$\kbr$>\ = \ 1.26\pm0.10$, $<$\kbi$> \ = \ 0.88\pm0.09$, where the errors
are standard deviations from mean. Typical 
measurement errors for \kbv\ are 0.2, and the measured $\beta$ values 
are quite consistent for all SN Ia. The small dispersions of the slopes 
are not a surprise and only confirm that Type Ia supernovae indeed form
a relatively uniform class of objects.

Although to first order the slopes $\beta$ are extinction independent,
they are weak functions of $K$-corrections which then implies that 
without applying $K$-corrections, $\beta$ will be a function of 
supernova redshift. The $K$-corrected fits of the $B$ versus $B-V$ color 
curves show slopes that are 
consistently smaller than those without $K$-correction. The $K$-corrected mean 
slope is \kbv\ = $1.94\pm0.16$. 

Histograms of the distribution of the slope \kbv\ are shown in Figure 7. To 
explore the potential of \kbv\ as an indicator of intrinsic properties of 
SN Ia, we have plotted the \kbv\ distribution in spirals and that in
E+S0 galaxies
separately. We have also constructed ideograms of the data by 
weighing each data point with associated errors (solid lines on Figure 7), 
and the expected 
distribution assuming a unique slope of 1.94 weighted by the fitted
slope errors of each supernova (dotted lines on Figure 7). For 
the supernovae in spiral galaxies,
the unique slope seems to provide an approximate description of the data. 
It is interesting to note, however, that the current data set is suggestive
of an intrinsic difference between the slope distribution for supernovae
in E+S0 galaxies and that in spirals. In particular, a hint 
of a secondary 
spike at \kbv\ around 1.65 is found for SN Ia from E+S0 but not 
for those
from spirals. Five out of 19 SN Ia with E+S0 hosts have \kbv\ 
around 1.65, whereas only two of 28 in spirals were found to have 
\kbv\ $\le$\ 1.65. The significance of these differences is hard to 
quantify considering the fact that the data collection
process may not be entirely homogeneous. For example, most of the SNe showing 
slopes around 1.65 were observed around 1992 (cf. Figure~9) which may be 
indicative of observational biases in the reported magnitudes. A 
Kolmogorov-Smirnov test gives a probability of 19\% for the two distributions
to be identical. This is not sufficient to firmly establish a statistical 
conclusion. Future observations should be able to resolve these issues.

\begin{figure}
\figurenum{7}
\epsscale{1.0}
\plotone{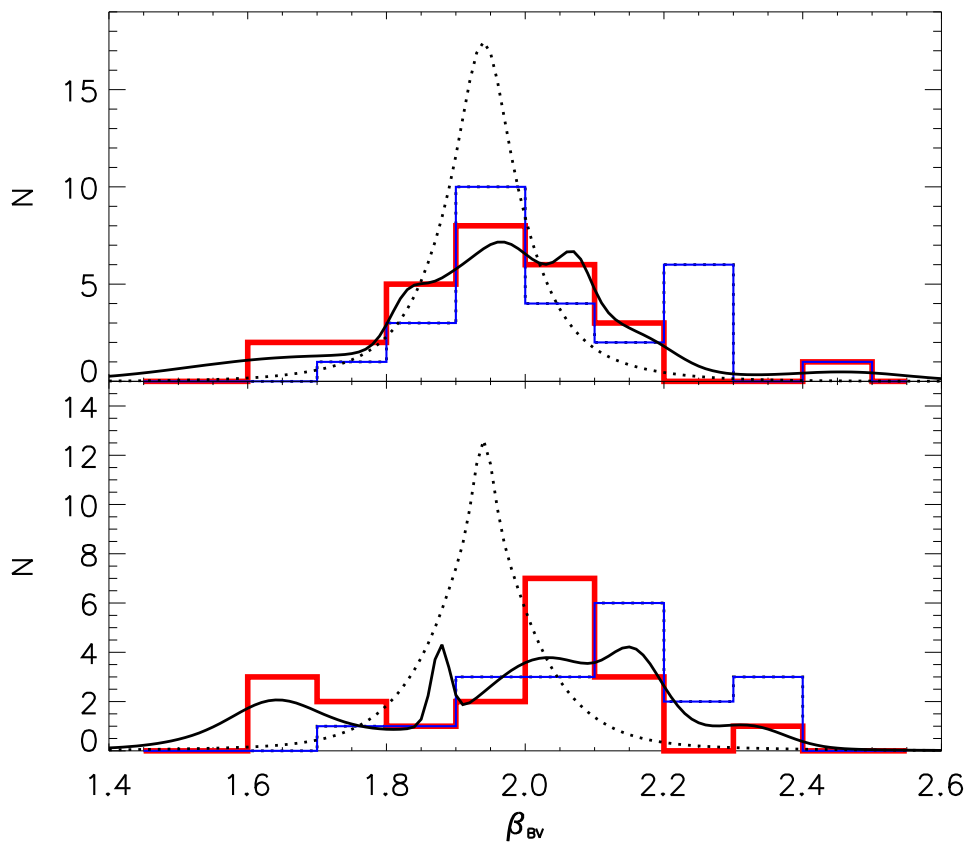}
\caption{Histograms of the distributions of the slopes of the linear fits.
The thin blue lines are without $K$-correction, and thick red lines are with 
$K$-correction. The smooth solid lines are the ideogram constructed by 
weighting each data point with associated errors. The dotted line shows
the distribution assuming a unique slope of 1.94 and then weighted by
the errors of each individual slope.
The upper panel is for supernovae in spirals and the lower for supernovae 
in E+S0 galaxies. More data would help to determine if 
the distribution in E+S0 galaxies are different from that in spirals.}
\end{figure}

\begin{figure}[h]
\figurenum{8}
\epsscale{1.0}
\plotone{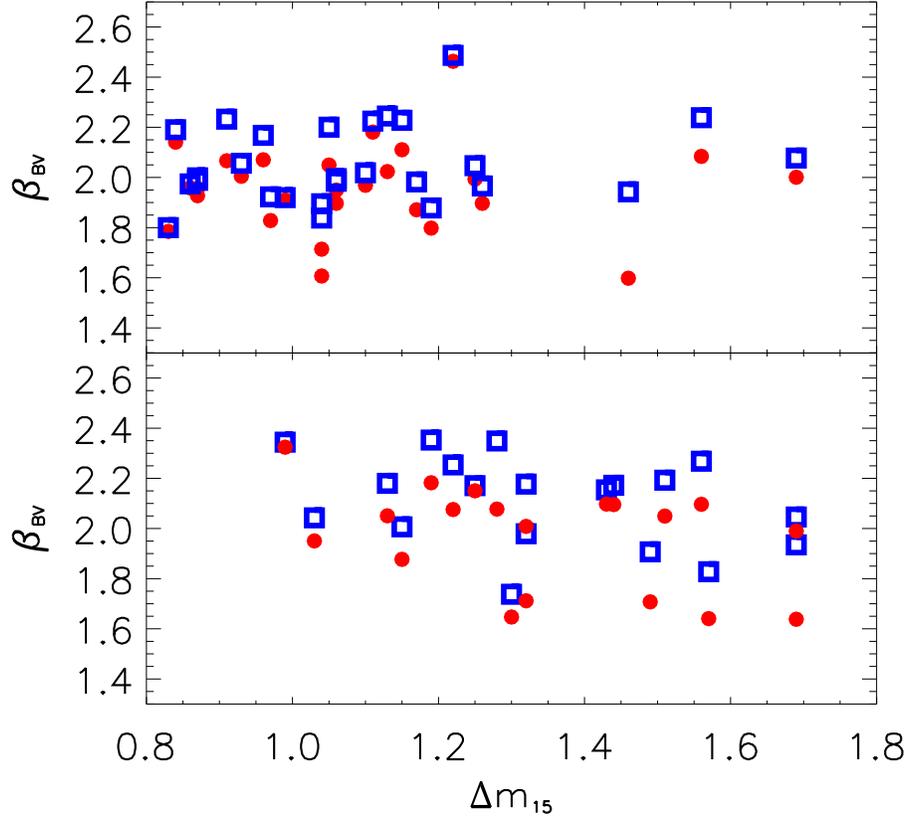}
\caption{The slopes of the Color-Magnitude fits versus the light curve
decline rate $\Delta m_{15}$ for supernovae from spiral(upper panel) and
E+S0 (lower panel) hosts. Red circles are for the slopes after
$K$-corrections, and blue squares are the data without $K$-corrections. 
No correlation is seen between the slopes and \dm15, as is expected 
from the definitions in equation 2.}
\end{figure}

\begin{figure}[h]
\figurenum{9}
\epsscale{1.0}
\plotone{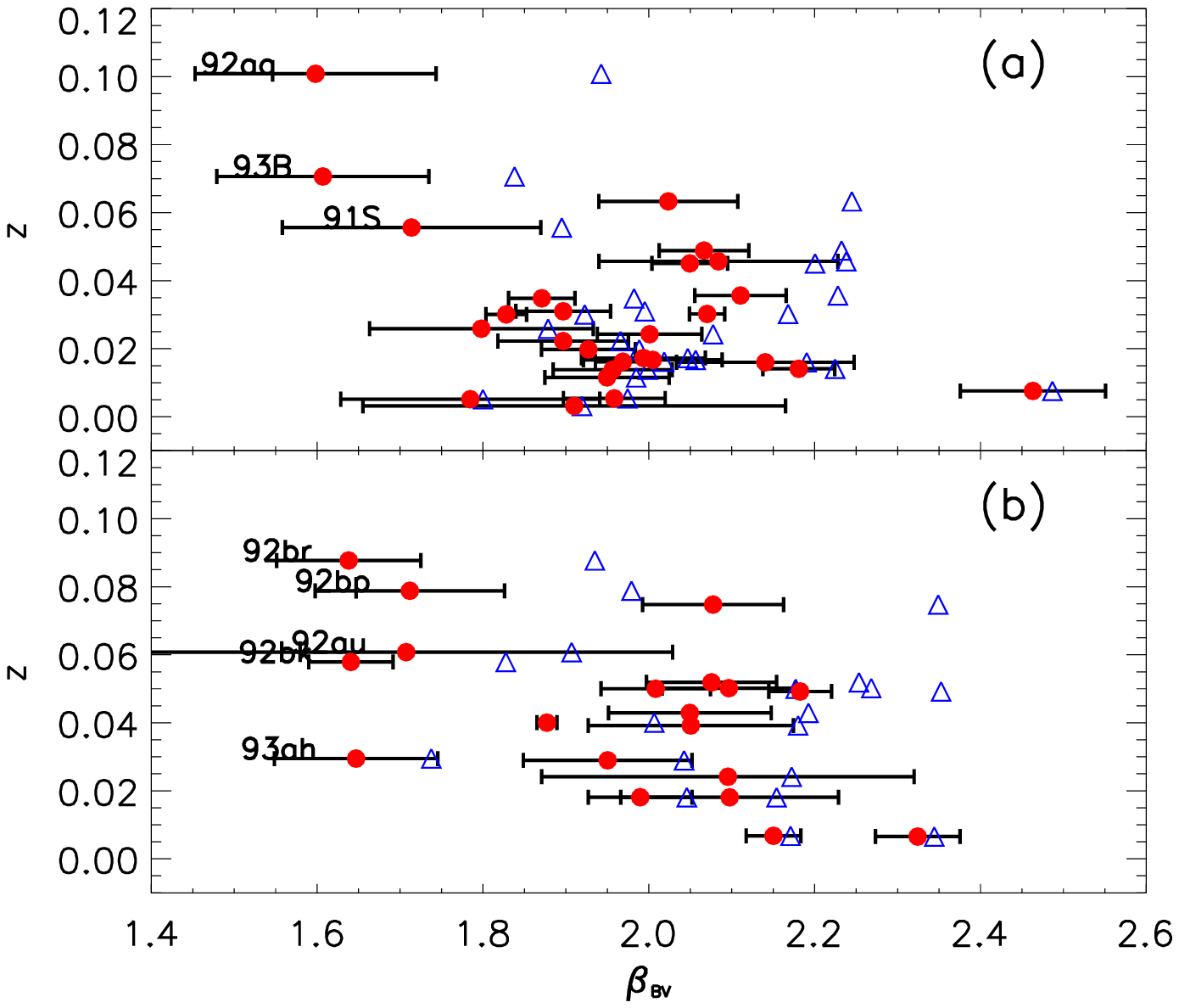}
\caption{Redshift versus slopes with (filled red circles ) and without 
(blue triangles) K-corrections for supernovae from spiral (a) and E+S0 
(b) galaxies. The secondary enhancement at \kbv\ = 1.65 shown in Figure 7(b) 
is also apparent in (b). The data are suggestive
of different branches of supernovae with intrinsically different slopes, but
more data are needed to test such differences.}
\end{figure}

It is interesting to note that $K$-correction does not appreciably decrease 
the \kbv\ dispersion. This might be an indication that we are indeed close 
to the limits of observational or $K$-correction errors. $K$-corrections are 
applied under the assumption that the supernovae under study are drawn from 
samples identical to those used to construct the $K$-corrections. If the 
assumption is correct, a reduction of slope dispersions is expected. 
In practice, the $K$-correction was constructed from a sample that is different 
in the mean from the larger sample we are studying, in a way such that some 
Type Ia with \kbv\ $\sim 1.65$ may be under-represented. A large 
number of spectroscopically well-observed supernovae are required to construct 
well calibrated, $K$-corrected slope distributions for different filters. 
Such distributions can be used for self-consistency checks of evolutionary 
effect of more distant supernovae. It may even be necessary to employ separate 
$K$-corrections for SNe with different \kbv\ values. Better $K$-corrections 
can be constructed in the future from systematic spectrophotometric 
supernova programs such as the 
Nearby Supernova Factory \citep{Aldering:2002a, Aldering:2002b}.

As we have already shown from discussions of equations (1) and (2), the 
slopes $\beta$ are uncorrelated or only weakly 
correlated to the decline rates or the stretch factors of the supernova 
light curves. This is easily understood from their definitions in 
equations~(2) where a stretch of both the $B$ and $V$ light curves does 
not affect the 
values of the slopes. The slopes change only when the {\it ratios} of the 
$B$ and $V$ stretches varies. And, indeed, as is shown in Figure 8, no 
obvious correlations of \kbv\ can be established with $\Delta m_{15}$, or 
the stretch factor.

As shown in the previous section and in Figure 6, 
the slopes of the linear fits are also functions of $K$-correction. For 
supernovae with stretch factor as the only characteristic parameter, applying 
$K$-corrections should remove the redshift dependence of \kbv. Figure~9~(a) 
and 9~(b) show \kbv\ and redshift correlation for SNe of spiral and 
of E+S0 hosts, respectively. The data suggest strongly that the 
assumption of a single parameter family is not appropriate in 
describing the variations of \kbv\ values. The possible \kbv\ enhancement  
around 1.65 revealed in Figure 7~(b) is seen again in Figure 9 
primarily for E+S0 hosts in Figure 9~(b). 
Comparing Figure 9~(a), and 9~(b) 
reveals that the incidence of SN Ia in E+S0 shows a possible deficit 
at \kbv\ around 1.85 compared to those in spirals. Note also that SNe below 
$z\ <\ 0.04$ are more tightly clustered than those at higher redshifts; this
is perhaps an indication of $K$-correction errors at higher redshifts.

\subsection{The Characteristics of \bbv}

\bbv\, \bbr, and \bbi\ by definition are characteristic magnitudes of the
linear CMAG fits evaluated at certain values of $B-V$, $B-R$, 
and $B-I$, respectively. These magnitudes can be used to derive distances to 
the supernovae just as the magnitude at optical maximum, which has been used 
historically.

\begin{figure}
\figurenum{10}
\epsscale{0.8}
\plotone{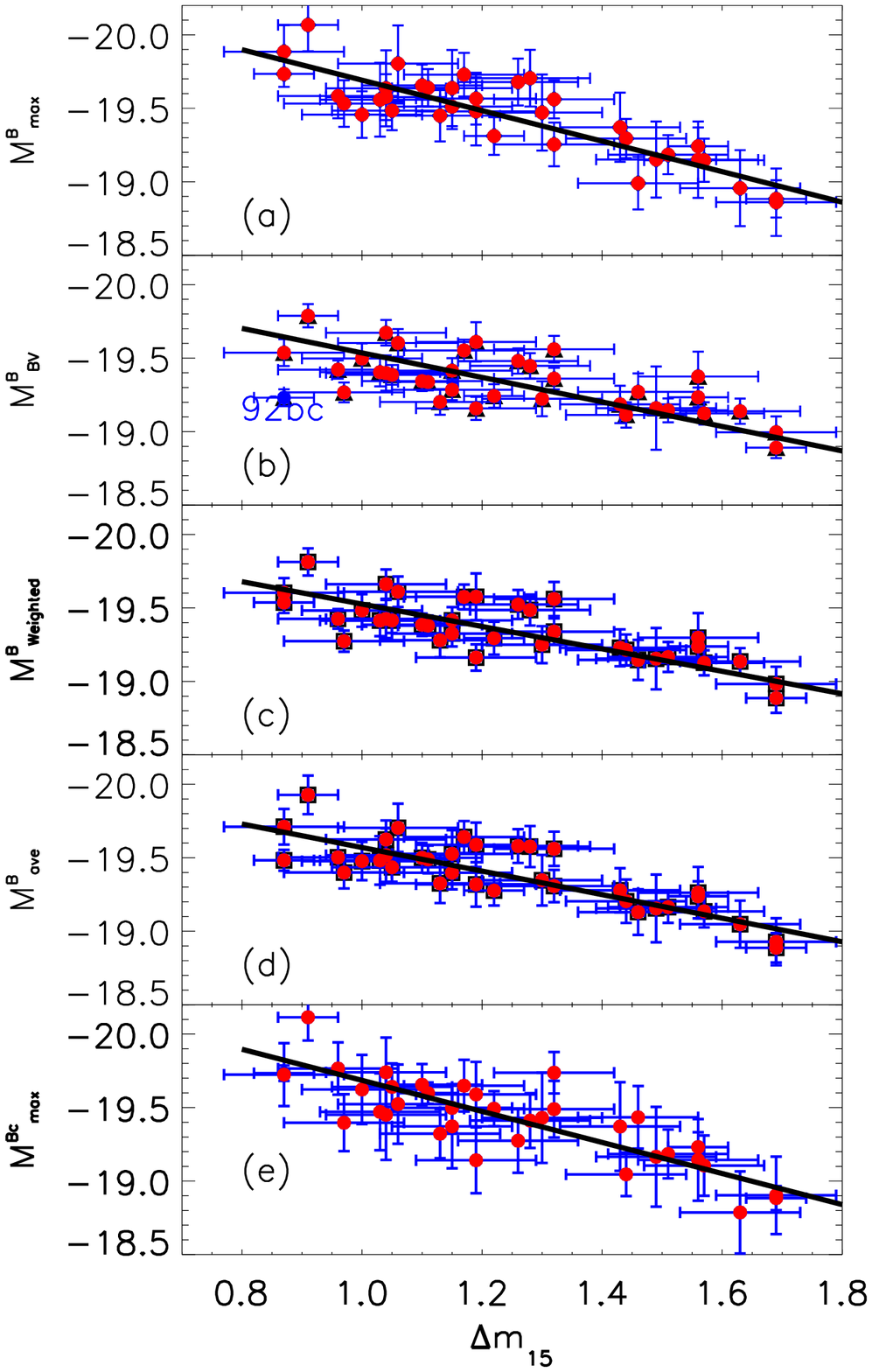}




\caption{Fits to the magnitudes and \dm15\ relations for $B_{max}$ (a),  
$B_{BV}$ (b), weighted average of the  $B_{max}$  and $B_{BV}$ (c),
unweighted average of $B_{max}$ and $B_{BV}$ (d), and $B_{max}$ (e) corrected
with extinctions derived from CMAGIC.  $R_B\ = \ 4.1$ in all fits and 
the constraint \BVm\ $<$ 0.2 is set to select the data points. In (b) SN 1992bc
is rejected from the fits at 3 $\sigma$ level. The solid lines show results 
from linear fits to the magnitude-\dm15\ relations. The average 
and weighted average magnitude show moderate improvement over both 
the original maximum magnitudes and the CMAGs. The magnitude dispersions after
the \dm15\ fits are shown in Table 2.}
\end{figure}

The most remarkable property of these color 
magnitudes is that extinction corrections to these magnitudes are not given 
by the conventional $A_B\ = \ R_B \cdot E(B-V)$. Instead, because of the color 
dependency of the magnitudes, extinction corrections are given by 
$$ A_{BV}\ = \ (R_B-\beta_{BV})\cdot E(B-V), \eqno(3a) $$
$$ A_{BR}\ = \ R_B \cdot E(B-V)\ - \ \beta_{BR} \cdot E(B-R), \eqno(3b) $$
and
$$ A_{BI}\ = \ R_B \cdot E(B-V)\ - \ \beta_{BI} \cdot E(B-I), \eqno(3c) $$
respectively.

With typical values of 
$R_B\ = \ 4.1$, $R_V\ =\ 3.1$, $R_R\ = \ 2.33$, and $R_I\ =\ 1.48$ and an 
extinction curve that is typical of the Galaxy \citep{Cardelli:1989},
we then find
$$ A_{BV}\ = (R_B\ - \beta_{BV}) \cdot E(B-V), \eqno(4a) $$
$$ A_{BR}\ = (R_B\ - \ 0.586 R_V \beta_{BR}) \cdot E(B-V), \eqno(4b) $$
and
$$ A_{BI}\ = (R_B\ - \ 0.858 R_V \beta_{BI}) \cdot E(B-V), \eqno(4c) $$
 
>From Table 1 where we found \kbv\ $\approx$ \ 2, we see that 
the extinction correction 
for \bbv\ is typically reduced by nearly 50\% of the extinction corrections 
needed for the original observed magnitudes. 
The CMAGs are therefore significantly less sensitive to errors in E(B-V) 
estimates. This 
is important because in the stretch or $\Delta m_{15}$ corrected magnitudes, 
errors in $E(B-V)$ are multiplied by about $R_B\ \approx\ 4.1$ and could 
easily become the dominate errors for distance measurements. A 50\% reduction 
of this error is a significant improvement. 

\begin{figure}
\figurenum{11}
\epsscale{1.0}
\plotone{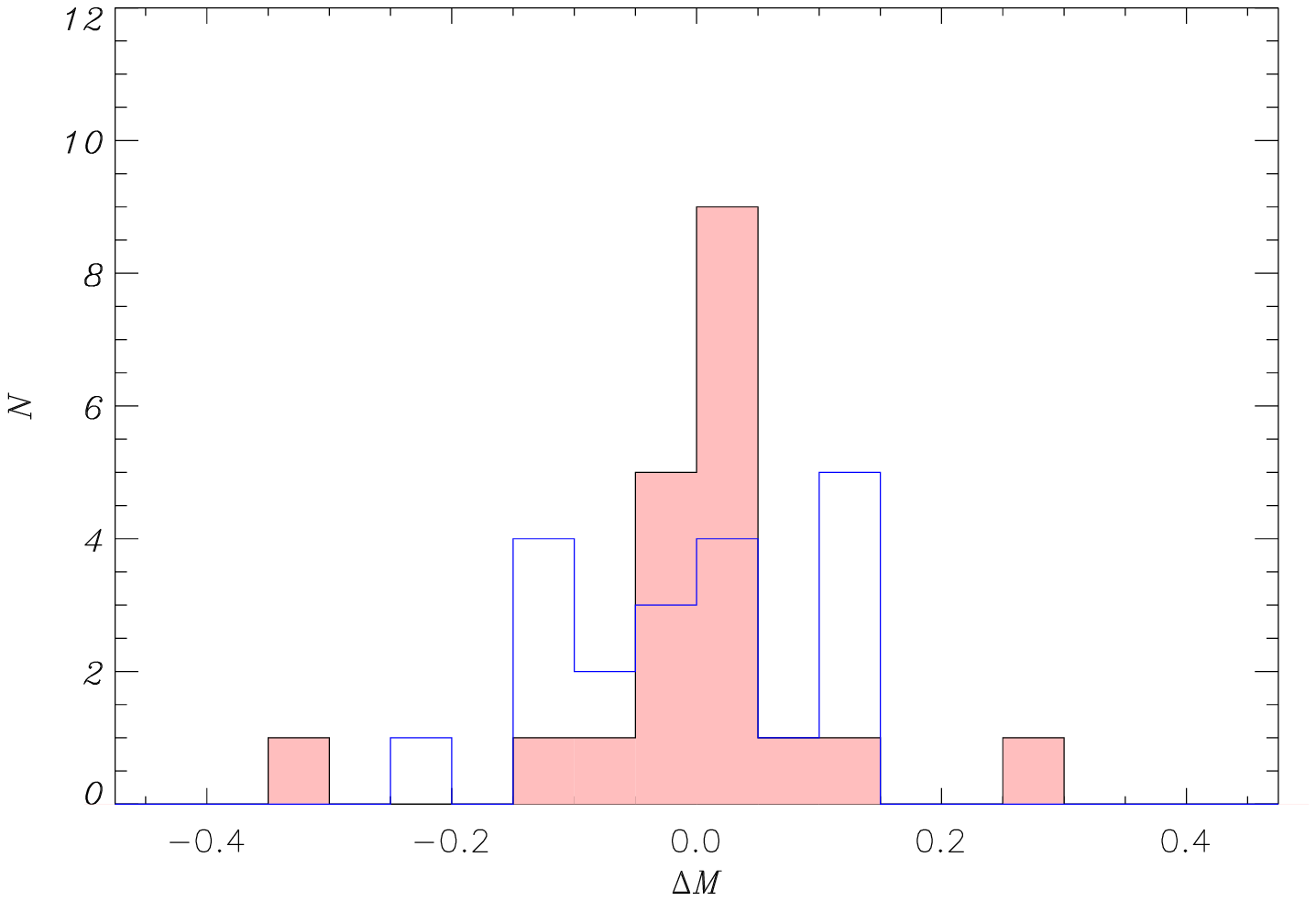}
\caption{Histograms showing the distribution of the residuals after
removing the \dm15\ dependence for a sample of low extinction 
supernovae with \BVm\ $<\ 0^m.05$. Interstellar extinction by the
host galaxy is removed assuming $R_B\ =\ 3.3$ which gives minimal 
magnitude dispersion. Apart from two 
apparent outliers (SN 1992bc and SN 1992bp), the Color-Magnitudes (pink 
shadowed line)
show significantly smaller dispersion than that of the maximum 
magnitudes (blue
solid line). Such improvement becomes less obvious for a larger
sample with more highly extincted supernovae included (cf lowest section of
Tables 2 and 3),
but this tight dispersion may be hard to see given extinction correction 
uncertainties.
}
\end{figure}

In the analyses that follow, 
we will only study the supernovae with redshift $z\ >\ 0.01$. 
To plot the various magnitudes on the Hubble diagram, we assume a 
peculiar velocity for all the SNe of 250 km/sec, and apply an extinction 
correction based on the $E(B-V)$ derived by \citet{Phillips:1999}. 

With no \dm15\ or stretch correction, the mean and standard deviations 
of the absolute color magnitude \Mbbv\ for the sample of supernovae with
\BVm\ $<\ 0^m.05$ are found to be -19$^m$.26 and 0$^m$.16, respectively.
The corresponding numbers for \MBm, the absolute magnitude at 
maximum are -19$^m$.38 and 0$^m$.26. This shows that for poorly observed 
data set where it is impossible to derive reliable \dm15, the color 
magnitudes can provide better distance estimates than the magnitudes at 
maximum.

The magnitude derived above, just like the magnitude at maximum, shows
correlations with the light curve decline rate. 
Figures 10~(a) and 10~(b) show linear fits of the conventional maximum magnitudes and \Mbbv\ 
versus \dm15, respectively. The fits were performed for supernovae with 
redshift above 0.01 and \BVm\ $\le$ 0.2 mag which excluded supernovae 
with large dust extinction. The color-magnitudes are correlated with 
\dm15, as can be seen in Figure~10~(b), but the corrections for 
\dm15\ dependence are smaller as the slope of the \dm15\ dependence is 
less steep than that for the maximum magnitudes. Note that SN~1992bc, 
an SN with a bump on the CMAG plot, deviates more than 3 $\sigma$ from 
the fitted line 
for $B_{BV}$. This could be the combination of a few factors such as
intrinsic differences of the supernovae and errors in extinction estimates.

\begin{figure}
\figurenum{12}
\epsscale{1.0}
\plotone{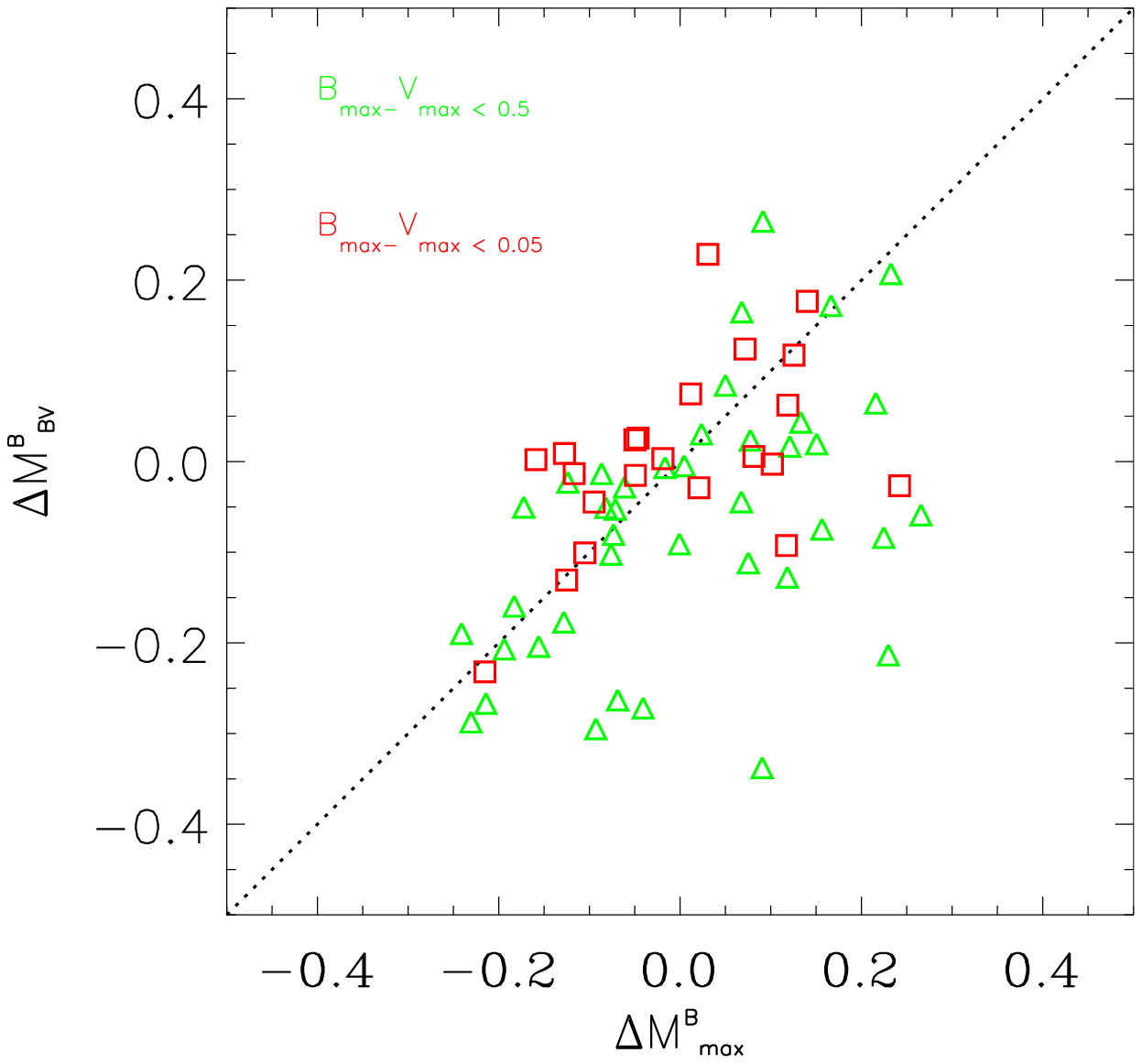}
\caption{Residuals of the magnitude-\dm15\ relation fits for \MBm\ and 
\Mbbv\ plotted against each other. The red squares are for the subsample of
supernovae with the most restrictive color cut \BVm\ $\le\ 0^m.05$, 
and the green triangles subsample includes less restrictive 
color cut \BVm\ $\le\ 0^m.05$.
Note that the supernovae that are common to 
both samples appear at different locations on this plot since the \dm15\
relations are different for the two samples.
Significant differences between the two magnitudes ({\it i.~e.}, points far from
the 45 degree line) are observed only for those 
with large extinctions.
}
\end{figure}

\begin{deluxetable}{l|l|l|l|l|l|l|l|l|l|l|l|l|l}
\tablecolumns{13} 
\tablewidth{0pt}
\tablecaption{Fit Parameters for \dm15\ Corrections with $R_B\ = \ 4.1$}
\tablehead{ 
\colhead{} & \multicolumn{6}{c}{No Outlier Rejections} & \multicolumn{6}{c}{Rejecting 3-$\sigma$ Outliers} \\  
\cline{3-5} \cline{8-12} \\
\colhead{\bf $M^B$}    &\colhead {a}        & \colhead{$\sigma${a}}      &  \colhead{b} & 
\colhead{$\sigma${b}} & \colhead{$\sigma$} &\colhead{N}  & \colhead{a}        & \colhead{$\sigma${a}}    
&  \colhead{b} & \colhead{$\sigma${b}} & \colhead{$\sigma$} & \colhead{N}
} 
\startdata
(1) & (2) & (3) & (4) & (5) & (6) & (7) & (8) & (9) & (10) & (11) & (12) & (13) \\
\cutinhead{B$_{max}$-V$_{max}\ \le\ 0.05$ (most restrictive) }
     {max}        &-19.57  & 0.04  & 0.95  & 0.15  & {\bf 0.10}  &20  &-19.57  & 0.04  & 0.95  & 0.15  & {\bf 0.10}  &20  \\
      BV    &-19.33  & 0.02  & 0.46  & 0.08  & {\bf 0.12}  &20 &-19.37  & 0.03  & 0.60  & 0.10  & {\bf 0.10}  &19  \\
     {Weighted}   &-19.40  & 0.03  & 0.62  & 0.10  & {\bf 0.08}  &20 &-19.40  & 0.03  & 0.62  & 0.10  & {\bf 0.08}  &20  \\
{Ave}   &-19.44  & 0.03  & 0.68  & 0.11  & {\bf 0.09}  &20  &-19.44  & 0.03  & 0.68  & 0.11  & {\bf 0.09} &20  \\
{EF}  &-19.09  & 0.03  &-0.07  & 0.11  & {\bf  0.16}  &20   &-19.09  & 0.03  &-0.07  & 0.11  & {\bf 0.16}  &20  \\
${max}^c$ &-19.58  & 0.04  & 1.01  & 0.12  & {\bf 0.12}  &20  &-19.58  & 0.04  & 1.01  & 0.12  & {\bf  0.12}  &20  \\

\cutinhead{B$_{max}$-V$_{max}\ \le\ 0.2$}
{max}        &-19.59  & 0.03  & 1.04  & 0.13  & {\bf 0.14}  &36 &-19.59  & 0.03  & 1.04  & 0.13  & {\bf 0.14}  &36  \\
{BV}    &-19.41  & 0.02  & 0.69  & 0.08  & {\bf 0.16}  &36  &-19.45  & 0.02  & 0.83  & 0.10  & {\bf 0.14}  &35  \\
{Weighted}   &-19.45  & 0.02  & 0.76  & 0.09  & {\bf 0.13}  &36  &-19.45  & 0.02  & 0.76  & 0.09  & {\bf 0.13}  &36  \\
{Ave}   &-19.49  & 0.03  & 0.80  & 0.10  & {\bf 0.13}  &36   &-19.49  & 0.03  & 0.80  & 0.10  & {\bf 0.13}  &36  \\
{EF}  &-19.17  & 0.02  & 0.14  & 0.11  & {\bf 0.21}  &36 &-19.17  & 0.02  & 0.14  & 0.11  & {\bf 0.21}  &36  \\
${max}^c$  &-19.58  & 0.03  & 1.06  & 0.12  & {\bf 0.14}  &36 &-19.58  & 0.03  & 1.06  & 0.12  & {\bf 0.14}  &36  \\

\cutinhead{B$_{max}$-V$_{max}\ \le\ 0.5$}
{max}        &-19.61  & 0.03  & 1.12  & 0.13  & {\bf 0.16}  &40 & -19.61  & 0.03  & 1.12  & 0.13  & {\bf 0.16}  &40  \\
{BV}    &-19.46  & 0.02  & 0.84  & 0.09  & {\bf 0.22}  &40  &-19.48  & 0.02  & 0.89  & 0.09  & {\bf 0.16}  &38  \\
{Weighted}   &-19.49  & 0.02  & 0.88  & 0.10  & {\bf 0.18}  &40  &-19.47  & 0.02  & 0.84  & 0.09  & {\bf 0.14}  &39  \\
{Ave}   &-19.53  & 0.03  & 0.89  & 0.10  & {\bf 0.15}  &40  &-19.51  & 0.03  & 0.87  & 0.10  & {\bf 0.13}  &39  \\
{EF}  &-19.18  & 0.02  & 0.04  & 0.12  & {\bf 0.24}  &40   &-19.18  & 0.02  & 0.04  & 0.12  & {\bf 0.24}  &40  \\
${max}^c$  &-19.58  & 0.03  & 1.02  & 0.11  & {\bf 0.15}  &40&-19.58  & 0.03  & 1.02  & 0.11  & {\bf 0.15}  &40  \\

\enddata 

\tablecomments{Fit results to the magnitude -- \dm15\ relations for $R_B\ =\ 4.1$ 
expressed in terms of ${\bf M^B}\ =\ (a\pm\sigma{a})+(b\pm\sigma b)\cdot\Delta m_{15}$, 
where {\bf $M^B$}represents
one of $M^B_{max}$, $M^B_{BV}$, $M^B_{Weighted}$, $M^B_{Ave}$, $M^B_{EF}$, or 
$M^{Bc}_{max}$. Columns
6 and 12 give the weighted deviations from the linear fits, and colums 7
and 13 give the number of supernovae used in the fits. 
}

\end{deluxetable}

\begin{deluxetable}{l|l|l|l|l|l|l|l|l|l|l|l|l|l}
\tablecolumns{13} 
\tablewidth{0pt}
\tablecaption{Fit Parameters for \dm15\ Corrections with $R_B\ =\ 3.3$}
\tablehead{ 
\colhead{} & \multicolumn{6}{c}{No Outlier Rejections} & \multicolumn{6}{c}{Rejeting 3-$\sigma$ Outliers} \\  
\cline{3-5} \cline{8-12} \\
\colhead{\bf M$^B$}    &\colhead {a}        & \colhead{$\sigma${a}}      &  \colhead{b} & 
\colhead{$\sigma${b}} & \colhead{$\sigma$} &\colhead{N}  & \colhead{a}        & \colhead{$\sigma${a}}    
&  \colhead{b} & \colhead{$\sigma${b}} & \colhead{$\sigma$} & \colhead{N}
} 
\startdata
(1) & (2) & (3) & (4) & (5) & (6) & (7) & (8) & (9) & (10) & (11) & (12) & (13) \\
\cutinhead{B$_{max}$-V$_{max}\ \le\ 0.05$ (most restrictive)}
max        &-19.55  & 0.04  & 0.98  & 0.13  & {\bf 0.10}  &20&-19.55  & 0.04  & 0.98  & 0.13  & {\bf 0.10}  &20  \\
{BV}    &-19.32  & 0.02  & 0.50  & 0.07  & {\bf 0.11}  &20 &-19.35  & 0.02  & 0.60  & 0.08  & {\bf 0.07}  &18  \\
Weighted   &-19.39  & 0.02  & 0.63  & 0.09  & {\bf 0.08}  &20&-19.39  & 0.02  & 0.63  & 0.09  & {\bf 0.08}  &20  \\
Ave   &-19.43  & 0.03  & 0.70  & 0.10  & {\bf 0.09}  &20&-19.43  & 0.03  & 0.70  & 0.10  & {\bf 0.09}  &20  \\
{EF}  &-19.13  & 0.02  & 0.14  & 0.09  & {\bf 0.13}  &20 &-19.13  & 0.02  & 0.14  & 0.09  & {\bf 0.13}  &20  \\
 ${max}^c$  &-19.57  & 0.03  & 1.03  & 0.11  & {\bf 0.11}  &20  &-19.57  & 0.03  & 1.03  & 0.11  & {\bf 0.11}  &20  \\

\cutinhead{B$_{max}$-V$_{max}\ \le\ 0.2$}
{max}        &-19.55  & 0.03  & 1.03  & 0.12  & {\bf 0.14}  &36 &-19.55  & 0.03  & 1.03  & 0.12  &{\bf  0.14}  &36  \\
{BV}    &-19.38  & 0.02  & 0.77  & 0.07  & {\bf 0.16}  &36&-19.38  & 0.02  & 0.72  & 0.08  & {\bf 0.11}  &32  \\
{Weighted}   &-19.42  & 0.02  & 0.81  & 0.08  & {\bf 0.13}  &36 &-19.43  & 0.02  & 0.80  & 0.08  & {\bf 0.12}  &35  \\
{Ave}   &-19.45  & 0.02  & 0.80  & 0.09  & {\bf 0.12}  &36 &-19.45  & 0.02  & 0.80  & 0.09  & {\bf 0.12}  &36  \\
{EF}  &-19.18  & 0.02  & 0.30  & 0.09  & {\bf 0.17}  &36 &-19.18  & 0.02  & 0.30  & 0.09  & {\bf 0.17} &36 \\
 ${max}^c$  &-19.56  & 0.03  & 1.06  & 0.11  & {\bf 0.14}  &36 &-19.56  & 0.03  & 1.06  & 0.11  & {\bf 0.14}  &36  \\
\cutinhead{B$_{max}$-V$_{max}\ \le\ 0.5$}
{max}        &-19.56  & 0.03  & 1.10  & 0.11  & {\bf 0.14}  &40 &-19.56  & 0.03  & 1.10  & 0.11  & {\bf 0.14}  &40  \\
{BV}    &-19.42  & 0.02  & 0.84  & 0.08  & {\bf 0.19}  &40 &-19.44  & 0.02  & 0.87  & 0.08  & {\bf 0.14}  &37  \\
{Weighted}   &-19.44  & 0.02  & 0.86  & 0.08  & {\bf 0.15}  &40 &-19.43  & 0.02  & 0.83  & 0.07  & {\bf 0.12}  &38  \\
{Ave}   &-19.47  & 0.02  & 0.85  & 0.09  & {\bf 0.13}  &40 &-19.46  & 0.02  & 0.84  & 0.09  & {\bf 0.12}  &39  \\
{EF}  &-19.20  & 0.02  & 0.26  & 0.09  & {\bf 0.19}  &40 &-19.20  & 0.02  & 0.26  & 0.09  & {\bf 0.19}  &40  \\
${max}^c$  &-19.55  & 0.03  & 1.03  & 0.10  & {\bf 0.14}  &40 &-19.55  & 0.03  & 1.03  & 0.10  & {\bf 0.14}  &40  \\

\enddata 

\tablecomments{Fit results to the magnitude -- \dm15\ relations  for $R_B\ = \ 3.3$
expressed in terms
of ${\bf M^B}\ =\ (a\pm\sigma{a})+(b\pm\sigma b)\cdot\Delta m_{15}$, where {\bf $M^B$} represents
one of $M^B_{max}$, $M^B_{BV}$, $M^B_{Weighted}$, $M^B_{Ave}$, $M^B_{EF}$, or 
$M^{Bc}_{max}$. Columns
6 and 12 give the weighted deviations from the linear fits, and collums 7
and 13 give the number of supernovae used in the fits.
}

\end{deluxetable}

We have also performed similar fits for different values of $R_B$ and 
color cuts. The resulting fitting parameters are shown
in Tables 2 and 3 where we also show the results after rejecting 
data which deviate by more than 3 sigma from the linear fits. 
The fits of Tables 2, 3, and 4 were all performed on sub-samples 
with different cuts on color \BVm\ $<\ 0^m.05$, $0^m.2$, and $0^m.5$. 
\BVm\ is basically an extinction indicator but is also correlated 
with intrinsic brightness 
of a supernova. Although not a requirement of CMAGIC, this cut 
simultaneously rejects heavily reddened and intrinsically red supernovae above
the given threshold. Most entries in Tables 2 and 3 are identical, but 
significant differences in dispersion (in bold face type) are observed for 
the sample with \BVm\ $<\ 0^m.05$.

Since there is independent measurement error for the brightness near maximum
and for the brightness at $B-V \ \sim\  0^m.6$, these two measurements may
be combined to get a better measurement. We test this by constructing 
different combinations of the maximum and color magnitudes, 
$$B_{ave}\ = \ (B_{max}\ + \ B_{BV})/2, \eqno(5a) $$
or in general, 
$$B_{Weighted}\ = \ B_{max} w\ + \ B_{BV}(1-w), \eqno(5b) $$
where $w$ is a weighting function that can be different for different
supernovae. Figure 10~(c) shows fits to $M^B_{Weighted}$, the absolute 
magnitudes derived from the weighted average of the magnitudes 
at maximum and the color magnitudes as given in equation (5b), 
with the weighting function chosen to be proportional to the inverse 
of the associated errors of $B_{max}$ and \bbv\ squared. Figure 10~(d) 
shows the fits to the absolute magnitudes of the unweighted average,
$M^B_{ave}$, for the average given in equation (5a).
A marginal improvement is seen when averaging the maximum 
and color magnitudes. However, the dispersions of the fits improve noticeably 
if we allow for rejection of a few outliers (cf Tables 2 and 3).

In Tables 2 and 3 
we show the fits for the absolute magnitudes at maximum and \Mbbv\
assuming two different values of $R_B \ = \ 4.1$ and $3.3$. The latter 
value is considerably smaller than the commonly 
adopted 4.1 of the Galactic extinction but seems to yields smaller dispersions 
for the $E(B-V)$ derived by \citet{Phillips:1999}. The reduction
of dispersion for the CMAGIC fits can be dramatic in some
cases, especially when 
significantly extincted supernovae are included in the fits.  
We see from Table 2 and 3, the dispersions of the 
magnitude versus \dm15\ linear fits are comparable for 
\Mbbv\ and the \MBm, but after rejecting two
outliers that deviate more than $3\sigma$ from the fits, the low 
extinction sample shows much smaller dispersion 
for \Mbbv\ ($\sigma_{B_{BV}}\ \sim\ 0^m.07$)  
than for \MBm\  ($\sigma_{M^B_{max}}\ \sim\ 0^m.10$).  
Histograms of the distributions of the 
residuals from the linear magnitude versus \dm15\ relations are shown in Figure
11 for \MBm\ and \Mbbv. It is clear that apart from
the two obvious outliers (SN 1992bc and SN 1992bp), the distribution is
suggestive of no significant intrinsic dispersion for \Mbbv.
 
To further understand the differences between \MBm\ and \Mbbv, 
we plot in Figure 12
the residuals of the magnitude versus \dm15\ fits for \MBm\ and \Mbbv.  
The sample of low extinction supernovae (with \BVm \ $<$ \ 0$^m$.05) is 
compared with another sample including more highly reddened supernovae 
(with \BVm \ $<$ \ 0.5). Note that the supernovae that are common to 
both samples appear at different locations on this plot since the \dm15\
relations are different for the two samples.
It can be seen from Figure~12 that significant differences 
between the two residuals are observed
mostly for those SNe with large extinctions. This shows also that extinction
correction is indeed one of the dominant sources of errors.

In principle, we can also construct an E(B-V)-free magnitude combination,
$$B_{EF} \ = \ B_{max}- R_B (B_{max}-B_{BV})/\beta_{BV} , \eqno(6a)$$
or equivalently,
$$B_{EF} \ = \ B_{BV} - (R_B-\beta_{BV}) (B_{max}-B_{BV})/\beta_{BV}, \eqno(6b) $$
It is easy to prove that the above equation is independent of $E(B-V)$. The
only extinction sensitive parameter is $R_B$. Because it involves
differences of the two magnitudes, this procedure usually introduces 
larger measurement errors when applying \dm15\ corrections 
(as shown in Tables 2 and 3)
but it has the advantage of being insensitive to the systematics of 
extinction determinations. Note also that, as shown in Tables 2 and 3, 
$B_{EF}$ requires significantly smaller \dm15\ corrections. The E(B-V)-free
estimator is useful when the supernovae are poorly observed and when
E(B-V) or \dm15\ measurements are difficult.

\section{Magnitude with Fixed slope $\beta_{BV}$}

We have shown that $\beta$ is a parameter with mean value 
$<$\kbv$> = 1.94\pm0.16$. Although the current data do not allow for a 
definitive conclusion on the distribution of the slope values, 
the small dispersion is remarkable.
During the epoch where the linear CMAGIC applies, typical B-V color varies  
from 0.2 to 1.0 mag. By choosing the CMAG intercept to be at 
$B-V\ = \ 0.6$, 
we see that the errors incurred by assuming a universal slope of 
\kbv\ = 1.94 for an actual range
of $\beta$ from 1.78 to 2.10 and the observed color $(B-V)_{obs}$ 
is around $0.16\ \times\ |(B-V)_{obs}-0.6|$, 
which is $\sim$ 0$^m$.06 if measured on the bluest or reddest
ends of the linear regions. In practice, the errors are much smaller as
the observations can be scheduled to obtain data close to 
$(B-V)_{obs}\ =\ 0^m.6$. The errors are then $\sim$ 0$^m$.03 for 
typical observations made around $0^m.4\ <\ (B-V)_{obs}\ <\ 0^m.8$ 
(which corresponds approximately to 10 to 20 days after $B$ maximum).
In this section, we assume that $\beta \ = \ 1.94 $ is 
universal for all of the supernovae, and use the linear part of the 
Color-Magnitude Diagram to fit the color magnitude intercepts. 

\begin{deluxetable}{l|l|l|l|l|l|l|l|l|l|l|l|l|l}
\tablecolumns{13} 
\tablewidth{0pt}
\tablecaption{Fit Parameters for \dm15\ Corrections with $R_B$ = 3.3 and fixed \kbv}
\tablehead{ 
\colhead{} & \multicolumn{6}{c}{No Outlier Rejections} & \multicolumn{6}{c}{Rejecting 3-$\sigma$ Outliers} \\  
\cline{3-5} \cline{8-12} \\
\colhead{\bf M$^B$}    &\colhead {a}        & \colhead{$\sigma${a}}      &  \colhead{b} & 
\colhead{$\sigma${b}} & \colhead{$\sigma$} &\colhead{N}  & \colhead{a}        & \colhead{$\sigma${a}}    
&  \colhead{b} & \colhead{$\sigma${b}} & \colhead{$\sigma$} & \colhead{N}
} 
\startdata
(1) & (2) & (3) & (4) & (5) & (6) & (7) & (8) & (9) & (10) & (11) & (12) & (13) \\
\cutinhead{B$_{max}$-V$_{max}\ \le\ 0.05$}
max        &-19.55  & 0.04  & 0.98  & 0.13  & {\bf 0.10}  &20  &-19.55  & 0.04  & 0.98  & 0.13  & {\bf 0.10}  &20  \\
BV    &-19.33  & 0.02  & 0.53  & 0.07  & {\bf 0.12}  &20  &-19.38  & 0.03  & 0.74  & 0.09  & {\bf 0.08}  &18  \\
Weighted   &-19.41  & 0.02  & 0.67  & 0.09  & {\bf 0.08}  &20 &-19.41  & 0.02  & 0.67  & 0.09  & {\bf 0.08}  &20  \\
Ave   &-19.44  & 0.03  & 0.71  & 0.10  & {\bf 0.09}  &20 &-19.44  & 0.03  & 0.71  & 0.10  &{\bf  0.09}  &20  \\
EF  &-19.13  & 0.02  & 0.14  & 0.09  & {\bf 0.13}  &20 &-19.13  & 0.02  & 0.14  & 0.09  &{\bf  0.13}  &20  \\
${max}^c$  &-19.57  & 0.03  & 1.03  & 0.11  & {\bf 0.11}  &20 &-19.57  & 0.03  & 1.03  & 0.11  & {\bf 0.11}  &20  \\

\cutinhead{B$_{max}$-V$_{max}\ \le\ 0.2$}
max        &-19.55  & 0.03  & 1.03  & 0.12  &{\bf  0.14}  &36 &-19.55  & 0.03  & 1.03  & 0.12  & {\bf 0.14}  &36  \\
BV    &-19.39  & 0.02  & 0.74  & 0.08  &{\bf  0.16}  &36 &-19.41  & 0.02  & 0.80  & 0.09  &{\bf  0.12}  &33  \\
Weighted   &-19.43  & 0.02  & 0.83  & 0.08  &{\bf  0.13}  &36 &-19.44  & 0.02  & 0.82  & 0.08  & {\bf 0.12}  &35  \\
Ave   &-19.46  & 0.02  & 0.81  & 0.09  & {\bf 0.12}  &36 &-19.46  & 0.02  & 0.81  & 0.09  & {\bf 0.12}  &36  \\
EF  &-19.18  & 0.02  & 0.30  & 0.09  & {\bf 0.17}  &36 &-19.18  & 0.02  & 0.30  & 0.09  &{\bf  0.17}  &36  \\
${max}^c$  &-19.56  & 0.03  & 1.06  & 0.11  & {\bf 0.14}  &36 &-19.56  & 0.03  & 1.06  & 0.11  &{\bf  0.14}  &36  \\

\cutinhead{B$_{max}$-V$_{max}\ \le\ 0.5$}
max        &-19.56  & 0.03  & 1.10  & 0.11  & {\bf 0.14}  &40 &-19.56  & 0.03  & 1.10  & 0.11  & {\bf 0.14}  &40  \\
BV    &-19.42  & 0.02  & 0.83  & 0.08  & {\bf 0.19}  &40 &-19.45  & 0.02  & 0.92  & 0.08  & {\bf 0.13}  &36  \\
Weighted   &-19.46  & 0.02  & 0.89  & 0.08  & {\bf 0.15}  &40 &-19.45  & 0.02  & 0.86  & 0.08  &{\bf  0.12}  &38  \\
Ave   &-19.48  & 0.02  & 0.86  & 0.09  & {\bf 0.13}  &40 &-19.47  & 0.02  & 0.85  & 0.09  & {\bf 0.12}  &39  \\
EF  &-19.20  & 0.02  & 0.26  & 0.09  & {\bf 0.19}  &40 &-19.20  & 0.02  & 0.26  & 0.09  & {\bf 0.19}  &40  \\
${max}^c$  &-19.55  & 0.03  & 1.03  & 0.10  & {\bf 0.14}  &40  &-19.55  & 0.03  & 1.03  & 0.10  & {\bf 0.14}  &40  \\

\enddata 

\tablecomments{Fit results to the magnitude -- \dm15\ relations for $R_B\ =\ 3.3$ 
expressed in terms of ${\bf M^B}\ =\ (a\pm\sigma{a})+(b\pm\sigma b)\cdot\Delta m_{15}$, 
where {\bf M$^B$} represents
one of $M^B_{max}$, $M^B_{BV}$, $M^B_{Weighted}$, $M^B_{Ave}$, $M^B_{EF}$, or 
$M^{Bc}_{max}$. Collumns
6 and 12 give the weighted devaiations from the linear fits, and collums 7
and 13 give the number of supernovae used in the fits.
}

\end{deluxetable}

The resulting magnitudes are given in Table 1, column 4, and the corresponding 
fits to the \dm15\ relations are shown in Table 4 for $z \ > \ 0.01$ and 
$R_B\ = \ 3.3$. Two sets of fits are shown in the Table, one with all of the
selected SNe, the other with all SNe whose magnitudes deviate by more than 
3 $\sigma$ being rejected from the fits. It is clear that the fixed slope 
works remarkably well, and is in general comparable in quality to the
variable slope fits. Typical dispersions for the sample with \BVm\ $\le\ 0^m.05$
are around $0^m.09$ for the color magnitudes and the various combinations
of the color magnitudes and the magnitudes at maximum.

As shown in Table 1, the fixed slope method provides equally good estimates
of CMAGs. It is interesting to note that in many cases the errors of 
magnitudes from the fixed slope method are smaller or comparable to those 
with variable slopes. The reason is obviously due to the fact that the
intercept is more constrained in a fixed slope fit than in a variable
slope fit.

Equation (6) can also be applied to the fixed slope method by replacing 
$\beta_{BV}$ with a constant. The corresponding \dm15\ corrections are 
given in Table 4, where most entries are close to those in Table 3.

\section{E(B-V) from CMAGIC}

Note that the last term of the color-excess free quantity given in 
equation (6a) is identical to the extinction corrections to $B_{max}$
with a color-excess of 
$${\mathcal E}(B-V)\ = \ (B_{max}-B_{BV})/\beta_{BV}. \eqno(7) $$ 
It is clear that ${\mathcal E}(B-V)$ is generally dependent on \dm15\ since
both \MBm\ and \Mbbv\ are \dm15\ dependent. The correlation 
of ${\mathcal E}(B-V)$
with \dm15\ is derived for the sample of SNe with \BVm\ $<\ 0^m.05$. Each SN was
corrected for Galactic dust extinction. The host galaxy extinction was 
corrected using $E(B-V)$ of \citet{Phillips:1999}; we do not expect 
significant errors from these assumptions for this low extinction sample. 
Linear fits to this sample of SNe gives
$${\mathcal E}_0\ =\ (-0.118\pm0.013)+(0.249\pm0.043)(\Delta m_{15}-1.1), \eqno(8)$$
where ${\mathcal E}_0$ denotes the supernovae that are unlikely to suffer
significant amounts of dust extinction. The results are shown in Figure 13.

\begin{figure}
\figurenum{13}
\epsscale{1.0}
\plotone{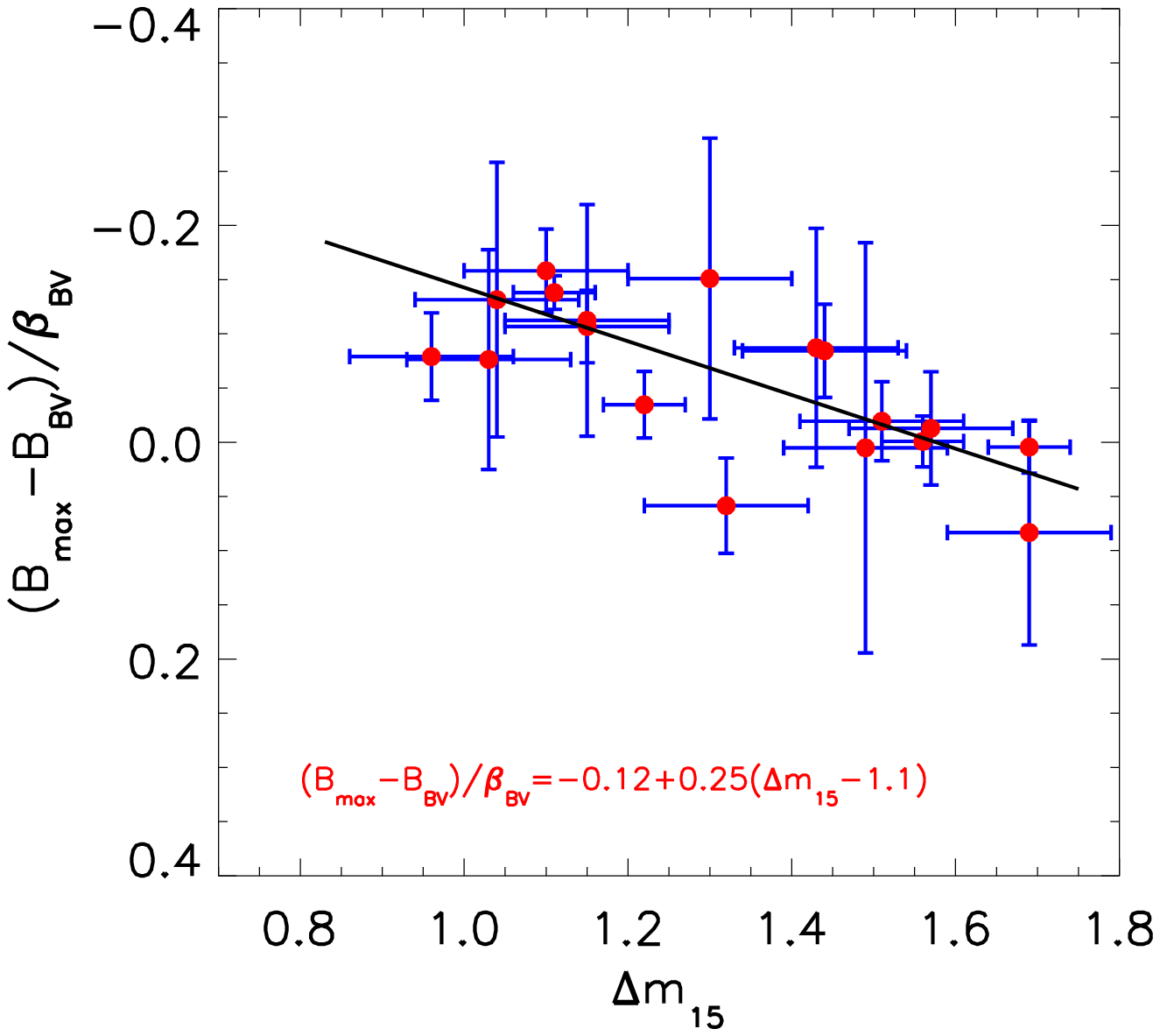}
\caption{Fits to the $(B_{max}-B_{BV})/\beta_{BV}$  versus \dm15\ relation for 
a sub-sample of SNe with \BVm\ $\le$ 0$^m$.05. The quantity 
$(B_{max}-B_{BV})/\beta_{BV}$ can 
be used to estimate the color excess due to extinction. }
\end{figure}

\begin{figure}
\figurenum{14}
\epsscale{1.5}
\plottwo{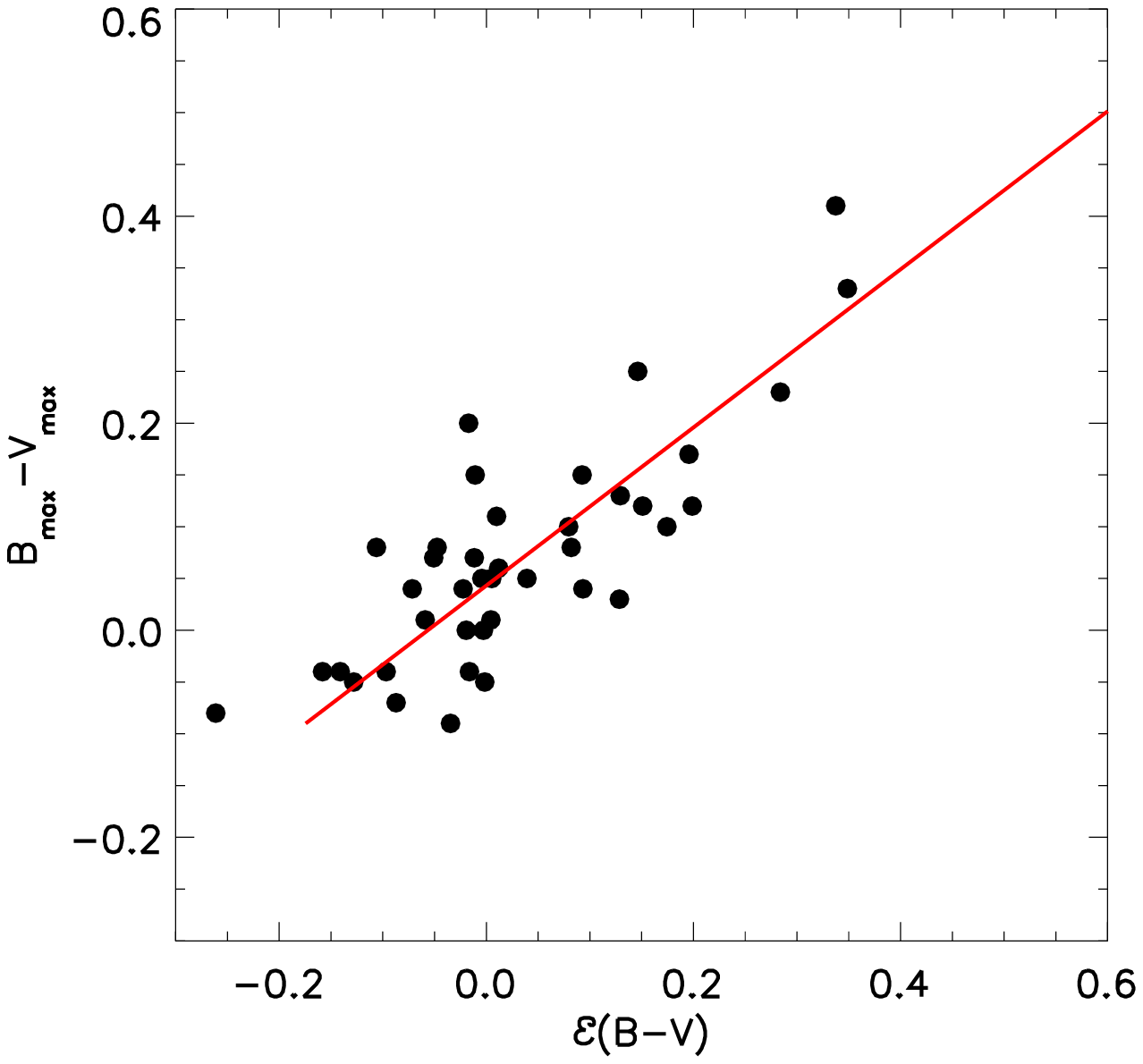}{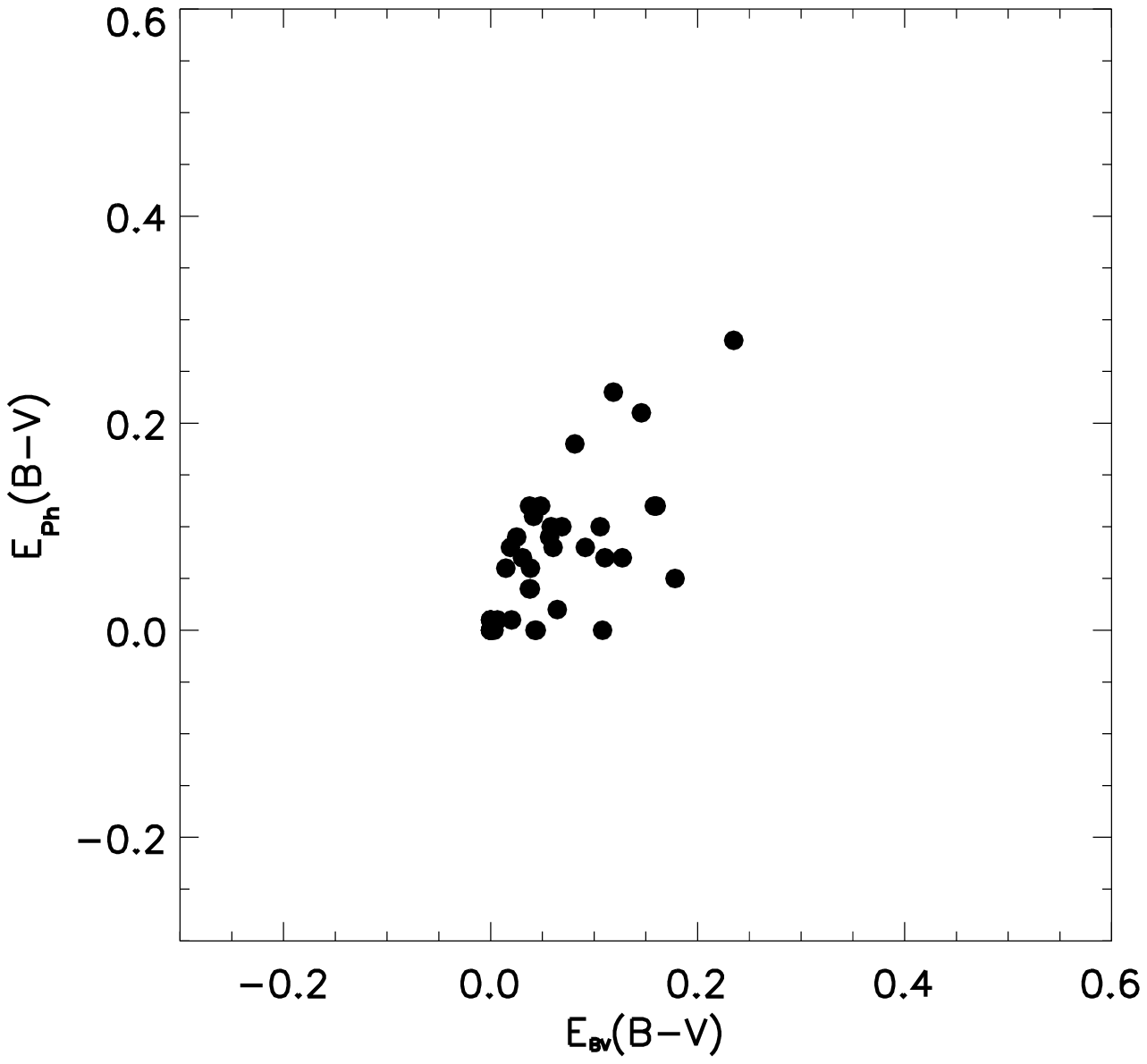}
\caption{Correlations of ${E}_{BV}(B-V) \ = \ (B_{max}-B_{BV})/\beta_{BV}$
with \BVm\ (a), and E$_{Ph}$(B-V) derived by Phillips et al (1999) (b). The
Bayesian Prior described in \citet{Phillips:1999} is used for both quantities
in (b).}
\end{figure}

Assuming that the above equation represents the \dm15\ dependence 
of ${\mathcal E}_0$ for supernovae with negligible  
extinction, we can derive the color-excesses of any SN Ia using the 
formula,

$${E}_{BV}(B-V) \ = {\mathcal E}(B-V)\ -\ {\mathcal E}_0. \eqno(9) $$

As a consistency check, we also analyzed the correlations between
${\mathcal E}(B-V)$ defined in equation (7) and
\BVm. The results are shown in Figure 14(a) for the supernovae 
with $B_{max}-V_{max}\ < \ 0.5$. The solid line in Figure 14(a) corresponds to
$${\mathcal E}(B-V)\ = \ (-0.051\pm 0.011)+(1.000\pm 0.067)(B_{max}-V_{max}). \eqno(10) $$
The relation given in equation (10) is interesting because the slope is 
consistent with 1. This means that ${\mathcal E}(B-V)$ is apparently 
a straightforward alternative to \BVm\ for measuring the color-excess.

To compare with the color-excess $E_{Ph}(B-V)$ 
deduced by \citet{Phillips:1999},
we have made use of equation (9) to calculate $E_{BV}(B-V)$ 
and applied  the same Bayesian prior used in \citet{Phillips:1999}.
The estimated color-excess values are shown in Column 9 of Table 1 and in Figure
14(b).
The corresponding correlation between the color excesses derived from 
${E}_{BV}(B-V)$ and those derived by \citet{Phillips:1999} is 
weaker than the one shown in Figure 14(a) but the values are 
nonetheless consistent. These are shown 
in Figure 14(b), where $E_{Ph}(B-V)$ denotes the color-excess derived
by \citet{Phillips:1999}. The Bayesian Prior is so strong here that the 
correlation in Figure 14(b) becomes less obvious.

It appears that the dispersion of the
magnitude at maximum after removing the \dm15\ dependence are comparable 
for the extinction corrections based on the color-excess derived from
equation (9) and for those based on the color-excess of \citet{Phillips:1999}.
The results of the \dm15\ fits are shown as $M^{Bc}_{max}$ in 
Tables 2, 3, and 4, which
should be compared with $M^B_{max}$ since they differ only by the extinction 
estimates. Note also that by definition, applying the color-excess derived from
equation (9) to \Bm\ and \bbv\ produces identical results.

It is interesting to compare the method outlined in equations 6, 7, 8, and 9 
with the method described in \citet{Tripp:1997} and \citet{Tripp:1999}, 
in which maximum magnitudes 
were fitted by \dm15\ and $B_{max}-V_{max}$ simultaneously to account for both 
intrinsic differences of SN luminosity and extinctions. The dependence on 
$B_{max}-V_{max}$ has a slope close to 2 in \citet{Tripp:1999}. This
slope is apparently related to both the intrinsic properties of a supernova
and the underlying extinction law which determines $R_B$.
\citet{Phillips:1999} found 
that a value around $R_B\ = \ 3.5$ provides better fits. In fact, we can 
also leave $R_B$ as a free parameter and make joint two parameter fits 
to \MBm\ versus \dm15\ and ${\mathcal E}(B-V)$. 
The current method will also be able to effectively separate the intrinsic 
color and the color-excess due to dust extinction, but we will
leave these treatments and self-consistent dust extinction estimates 
for a later paper.

\section{Observational Strategies for CMAGIC}

It is noteworthy that the CMAGIC method is applicable in cases where the SN 
is first observed well past maximum light, an epoch at which the extrapolated 
fit to the 
light curve maximum becomes questionable. It can also simplify SN Ia observations
for distance estimates. There are several strategies that may take advantage 
of this method, which we describe in order of increasing number of
measurements needed on each SN. 

First, in the extreme case, a single pair of data points with
two color photometry from a color-excess free supernova such as those
from E+S0 galaxies may be used to obtain a distance modulus 
estimate
that is accurate to about 0.16 mag even without knowledge of the stretch
factor if the co-moving frame $B$ and $V$ 
filters are used. 
As shown in \S 3, systematic errors introduced by assuming a 
universal $\beta_{BV}$ value are less than $\sim 0^m.06$ and more 
typically $\sim 0^m.03$ when $B-V$ is measured within $\pm0^m.2$ of 0$^m.6$.
The observations have to be made in the linear regime, typically around
day 10-20 so that CMAGIC is applicable. A rough estimate of the epoch and
redshift must be obtained independently most likely by using a single
spectrum \citep{riess:1998}. This simple method does not provide 
proper extinction estimates and can be applied only when extinction can be 
estimated independently. 
Such sparse data points may be useful for high redshift supernovae
where detailed light curves are hard to obtain. This method might be 
termed the ``CMAGIC-Snapshot'' method, but should be applied with caution
as there is no control of systematics. 

A significant improvement is achieved if, in addition to color 
information around day 10-20, a single-band
light curve is also obtained at a certain color that allows for a 
fit to the maximum magnitude and stretch factor as has been 
routinely done in current supernova observations. 
With such a data set, stretch/\dm15\ corrections can be applied,
and a color-excess-free estimator can be constructed. It is also practical
to estimate dust extinction using the method described above. Such data 
can be obtained for most of the Type~Ia supernovae discovered by modern 
techniques, from low redshift to redshift as high as $\sim\ 0.85$ from 
the ground, but 
space-based observatories are needed to observe supernovae at even higher 
redshifts in the multiple bands to get extinction 
\citep{Aldering:2002a, Aldering:2002b}.

If more color information and light curve information is available, joint
fits to stretch/\dm15\ correction and CMAGIC can be attempted to
reduce errors on the color-excess free estimator and even absolute extinction
$A_\lambda$ to achieve unbiased distances. In particular, we have shown that
the color magnitude relation is a very robust relation that none of
the current standard fitting procedures can match in quality. 
The Color Magnitudes are less sensitive to errors in extinction corrections. 
The corrections for light curve
shapes are smaller and therefore are less sensitive to errors of \dm15\ or 
stretch. These properties make CMAGIC a more robust and reliable 
method for analyzing SN Ia light curves.

\section{Discussions and Conclusions}

We have presented in this study a new way to analyze photometric data of SN Ia
for cosmological distance determinations. Our method is based on post-maximum
observations of SN Ia and we find that a precise distance estimate 
which corrects simultaneously for the $\Delta m_{15}$/stretch relation
and color-excess can be obtained with well-sampled light curves in
different filters. We have seen that the magnitude of Type Ia SNe at a given 
color index has a very small dispersion for most of the first month past 
maximum, and the linear fit in this region is simple and robust for 
$B$, $V$, $R$, and $I$ data. 
Working in the color-magnitude plane rather than magnitude- or 
color-versus-time captures better the homogeniety of SN Ia.
It also is less affected by errors in extinction estimates in most bands. Finally it allows a different part of the SN lightcurve to be explored in hunts for new evolutionary parameters. This method may be useful for ground- or space-based 
multi-color supernova surveys. 

CMAGIC emphasizes post maximum data and gives zero weight for data taken 
outside the linear regime. 
Compared with other methods, CMAGIC is more transparent and error 
propagation is more straight forward.
The fundamental difference between
the CMAGIC approach and other approaches such as 
stretch fit \citep{Perlmutter:1997},
template fit \citep{Pskovskii:1977, Phillips:1993}, and MLCS \citep{Riess:1996}
is that CMAGIC uses the color magnitude plane and particularly the linear relation found for all Type Ia supernovae; it 
does not employ any template light curves. The differences in input 
mathematical models determine that CMAGIC is intrinsically different
from MLCS. There is no reason to suppose that a linear sequence of
time parametrized light curves, such as in MLCS, is guaranteed to 
capture all variations of Type Ia SNe.
 In particular, as shown in Figure 6b, the fitting
templates of MLCS from Riess et al. (1996) can not reproduce the bump 
feature on the CMAG diagram which is observed for many intrinsically 
bright supernovae. (The magnitude dispersion on the Hubble Diagram given
in \citet{Riess:1996} is generally larger than what we derived from CMAGIC
for comparable samples of supernovae.)
Future versions of the MLCS, template fitting, or stretch fitting 
approaches may or may not be able to incorporate the 
CMAGIC empirical relation, but  the current implementations do not.

One important application of the method is to derive color-excesses
due to dust extinction. The method discussed in \citet{Phillips:1999} which
makes use of nebular phase data
is hard to apply to faint distant supernovae. The extinction 
estimate provided in this study can be a useful tool for both
high and low redshift supernovae. 

Fitting the slope of CMAGIC may come as a handy parameter to probe the
progenitor systems. The slope is an observable parameter and we have already seen 
from Figures~7 and 9 that there is an intrinsic dispersion of the slopes 
which can not be accounted for by observational errors. Figures~7 and 9 
are suggestive that we may have seen a bifurcation of the slope 
distribution with one branch of SN Ia showing slopes of about 
$\beta_{BV}\ \sim\ 1.65$ and the other showing about $\beta_{BV}\ \sim\  2.0$. 
If this is real, it implies that we may be witnessing some fundamentally 
different groups of supernovae with different chemical structures or evolutionary paths. 
These questions will be extremely interesting questions to be
addressed by systematic supernova observation programs.

Detailed radiative transfer calculations are needed to understand the
CMAGIC relations. A possible key factor for the success of CMAGIC
is that the ejecta of Type Ia supernovae are more
uniform in the zones of complete nuclear burning which are only visible
after maximum light. The CMAGIC relations are determined mostly by the 
chemical and ionization structures close to the photosphere
which for SNe Ia are perhaps more uniform as they lie in regions 
of complete thermal nuclear reaction \citep{Hoeflich:1998}. 

One other line of evidence for this homogeneity just past maximum
light comes from polarization studies.
Spectropolarimetry of SNe Ia indicate that the supernova photospheres
are more asymmetric before and around optical maximum 
\citep{Wang:1996, Howell:2001, Wang:2002}. 
The value representing the asphericity of the photosphere 
can be determined to be around 10\%, which would introduce dispersions due
to view-angle differences of up to 0.1 magnitude before and around 
optical maximum, but most post maximum observations show 
non-detectable polarizations down to about 0.2-0.3\% 
\citep{Wang:1996, Wang:2002}, indicating that
their post maximum behavior can be more uniform than before optical 
maximum. 

The dramatic improvement in magnitude dispersion 
as implied in Figure 11, if borne out by
future nearby supernova observations, would have important 
impact for the use of Type Ia supernovae as cosmological 
distance indicators.

\acknowledgments This work was supported by the Physics Division, E. O.
Lawrence Berkeley National Laboratory of the U. S. Department of Energy
under Contract No. DE-AC03-76SF000098. LW is grateful to Zongwei Li
of Beijing Normal University and Jianshen Chen of Beijing University for
a visit to China where part of the work was performed.

\clearpage

\appendix

\section{The Complete Set of Magnitude versus Color plots}

We show in Figures A1, A2, and A3 the color-magnitude plots for 
B-V, B-R, and B-I colors, respectively.
All these Figures represent the magnitudes and colors as measured with no
extinction or $K$-corrections. Linear regimes can be identified for all the
colors.

\begin{figure}[h]
\figurenum{A1}
\epsscale{1.0}
\plotone{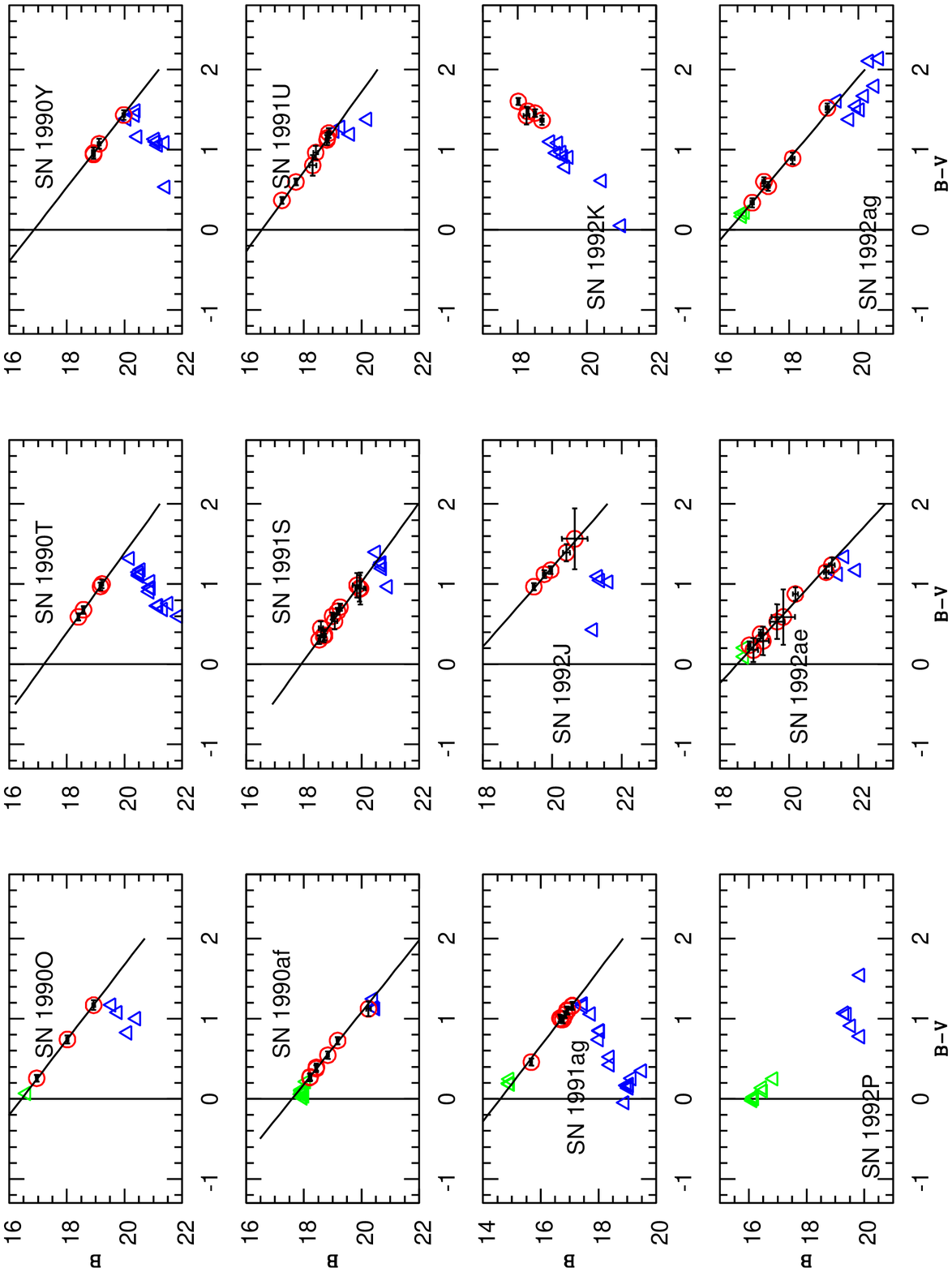}
\end{figure}

\begin{figure}
\figurenum{A1}
\epsscale{1.0}
\plotone{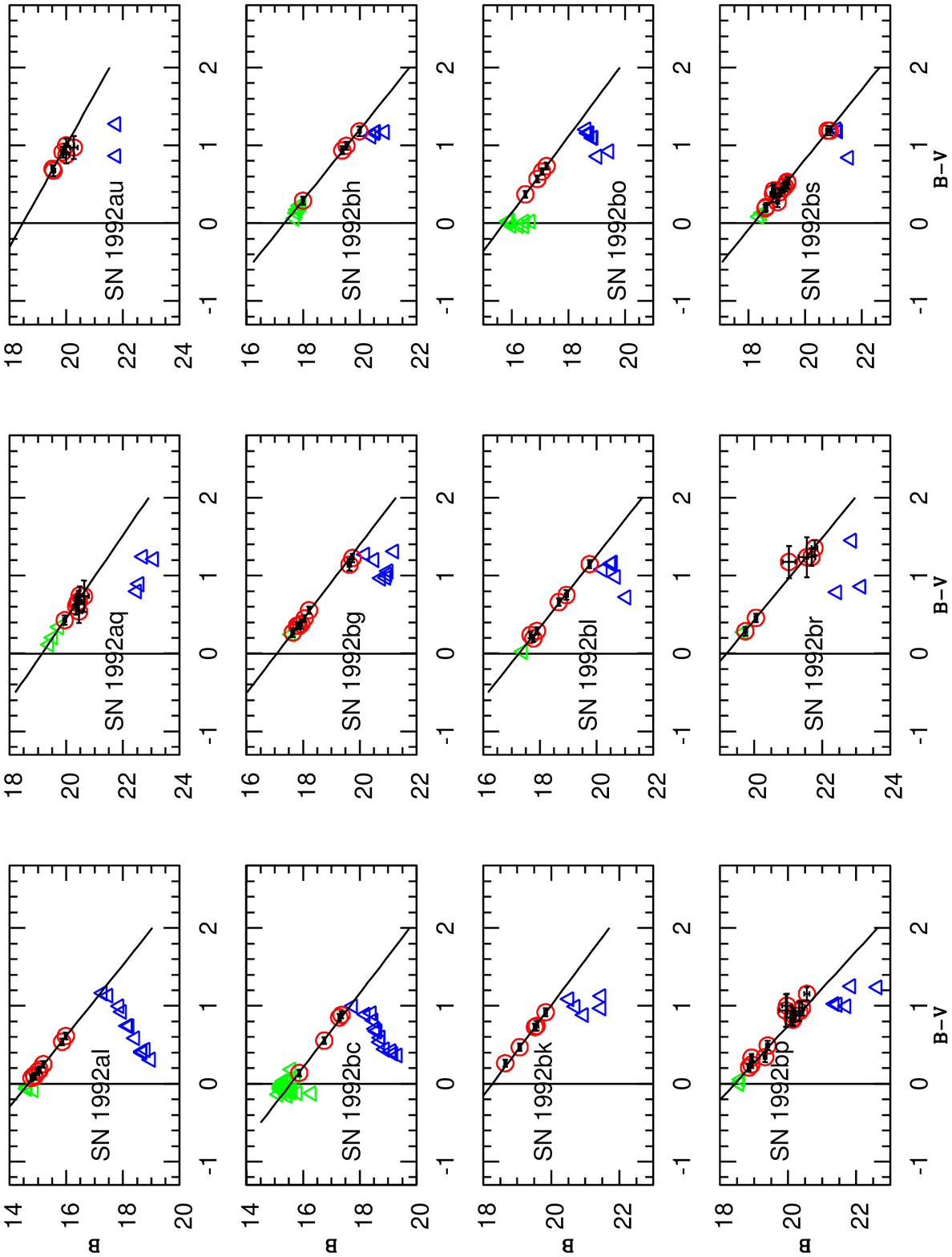}
\end{figure}

\begin{figure}
\figurenum{A1}
\epsscale{1.0}
\plotone{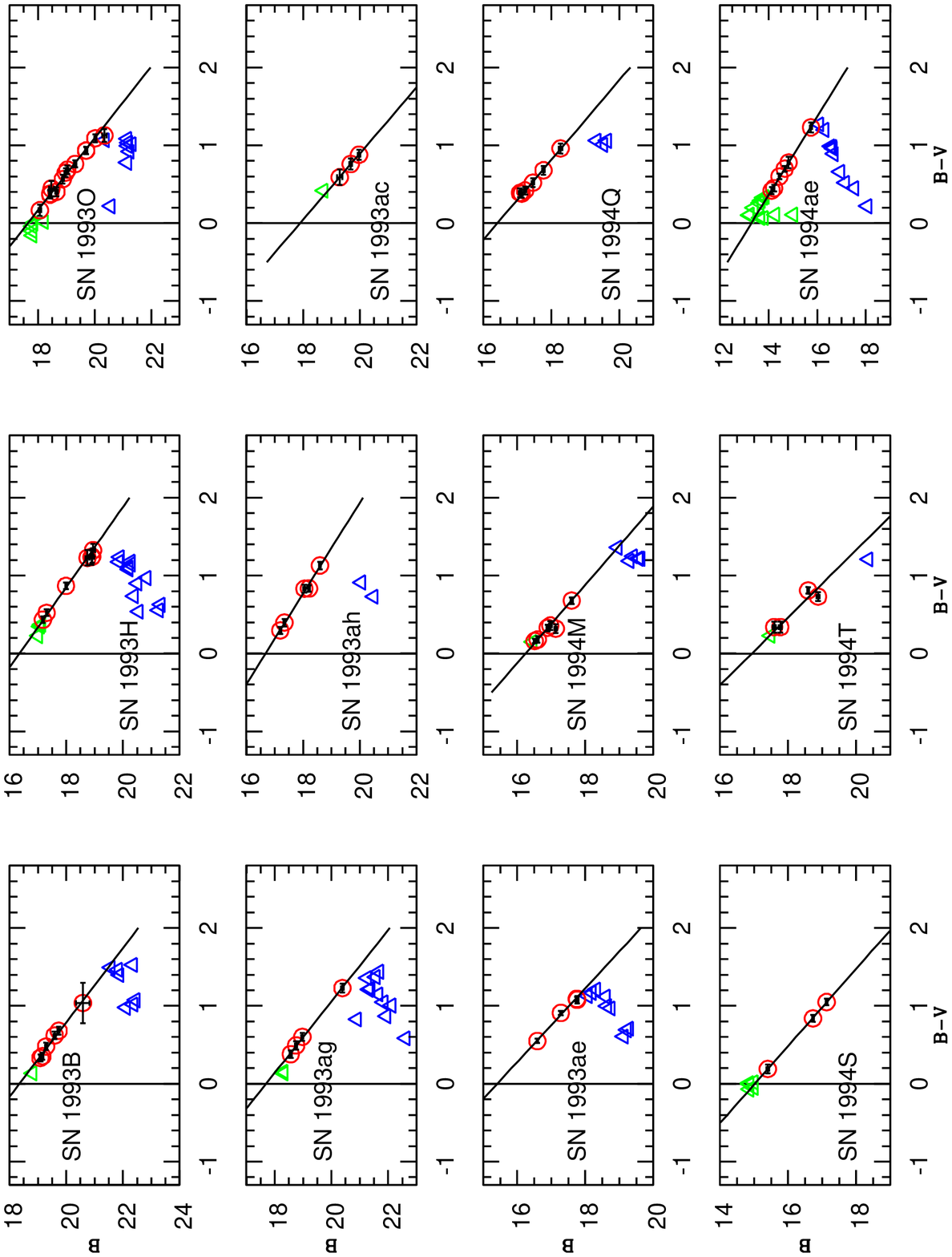}
\end{figure}

\begin{figure}
\figurenum{A1}
\epsscale{1.0}
\plotone{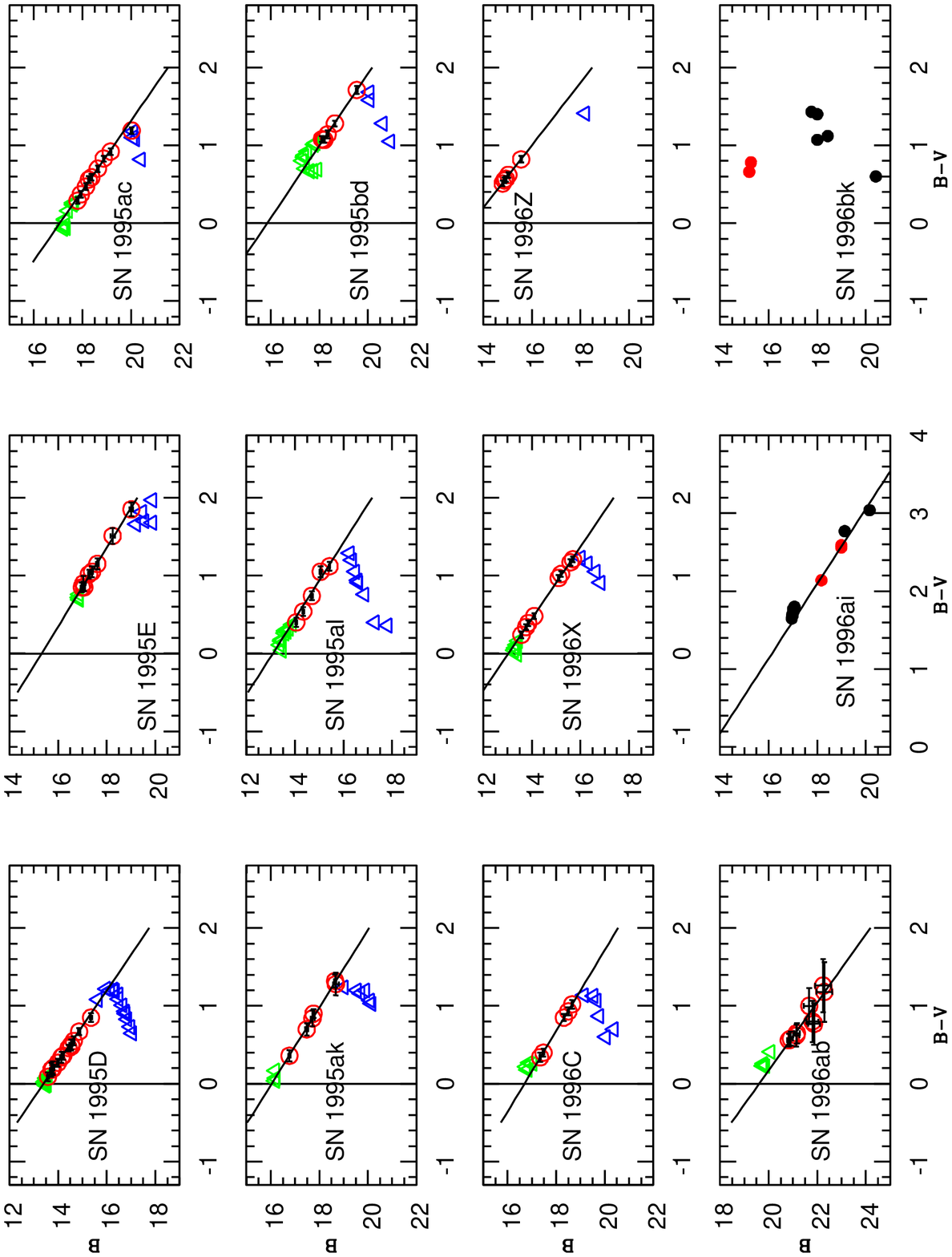}
\end{figure}

\begin{figure}
\figurenum{A1}
\epsscale{1.0}
\plotone{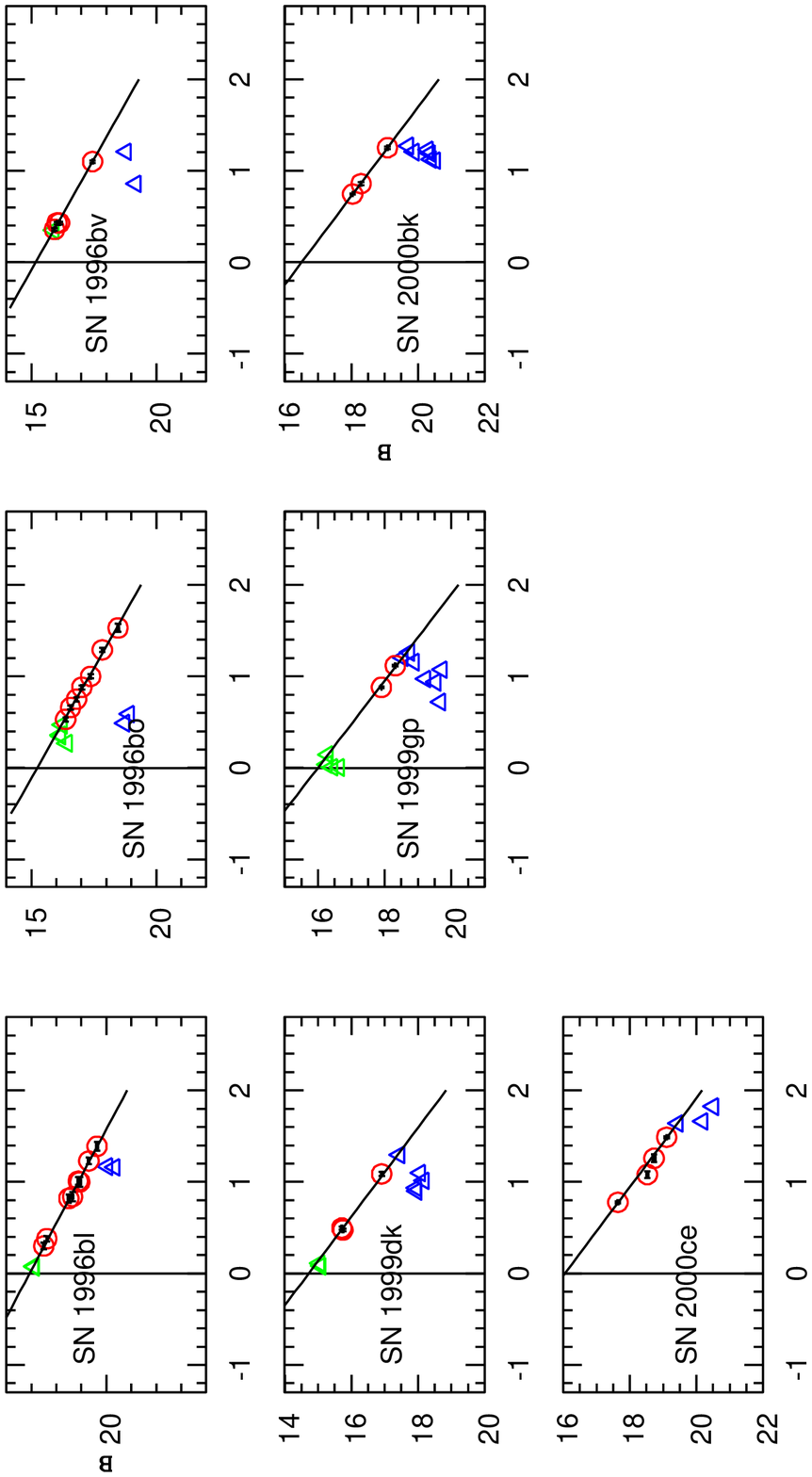}

\caption{Observed B magnitude versus B-V for the sample of well
observed SNe ({Hamuy et al. 1996}; {Riess et al. 1999}; 
Krisciunas et al. 2001).  The linear fits are
shown in solid lines. The data points used for the
linear fits are shown as red circles. The green triangles show data points
before the linear region, and the blue triangles show data after the linear 
region.}
\end{figure}

\begin{figure}
\figurenum{A2}
\epsscale{1.0}
\plotone{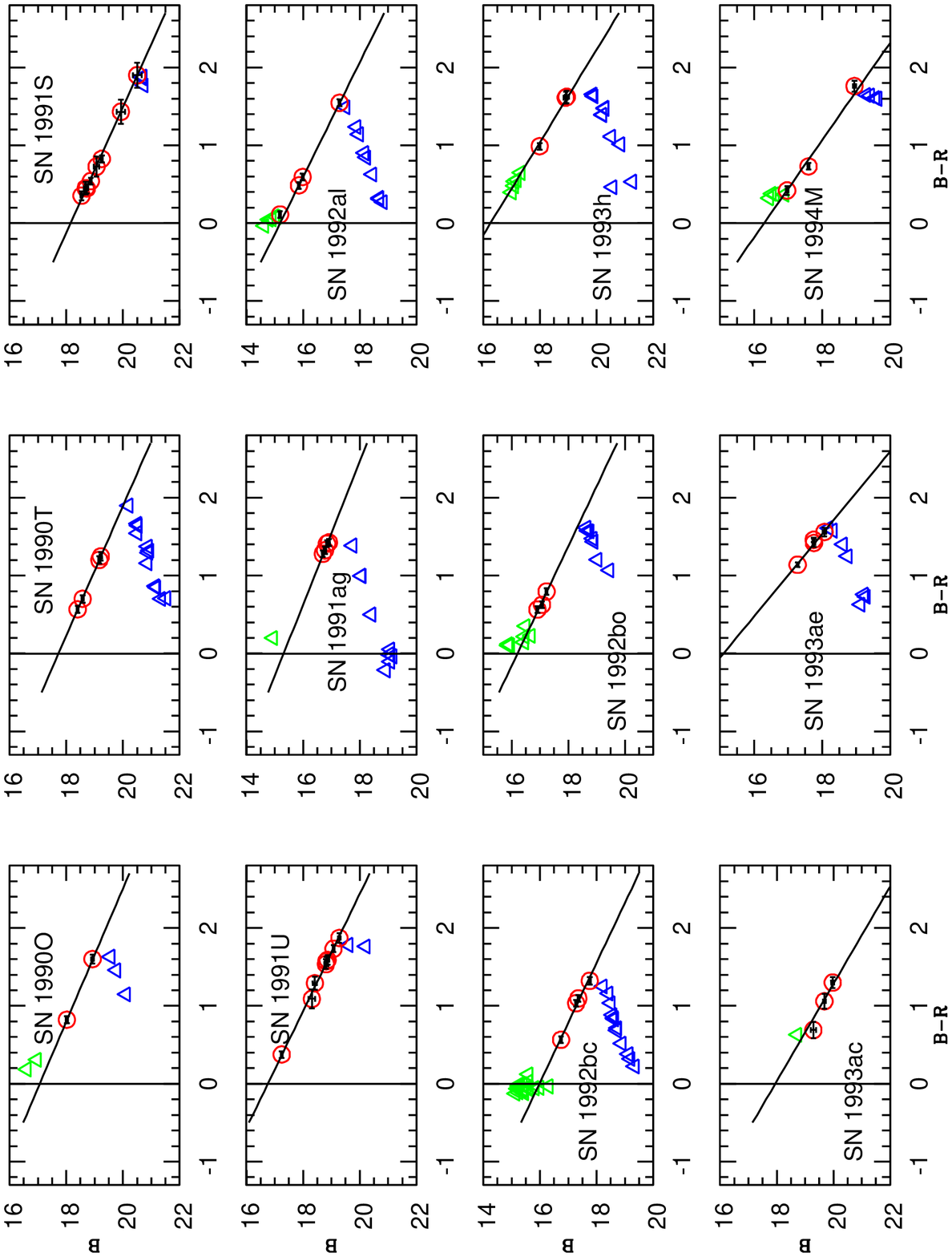}
\end{figure}

\begin{figure}
\figurenum{A2}
\epsscale{1.0}
\plotone{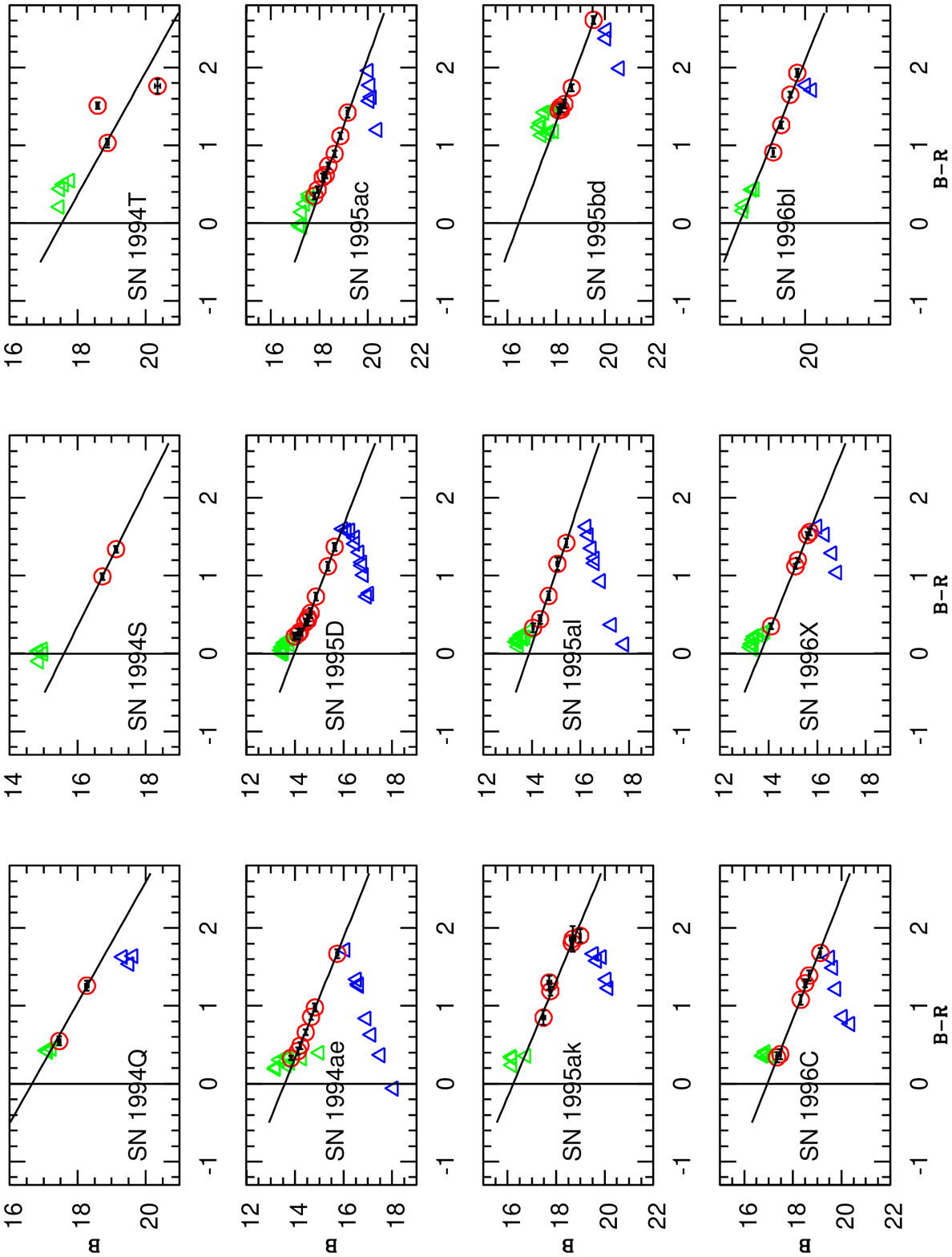}
\end{figure}

\begin{figure}
\figurenum{A2}
\epsscale{1.0}
\plotone{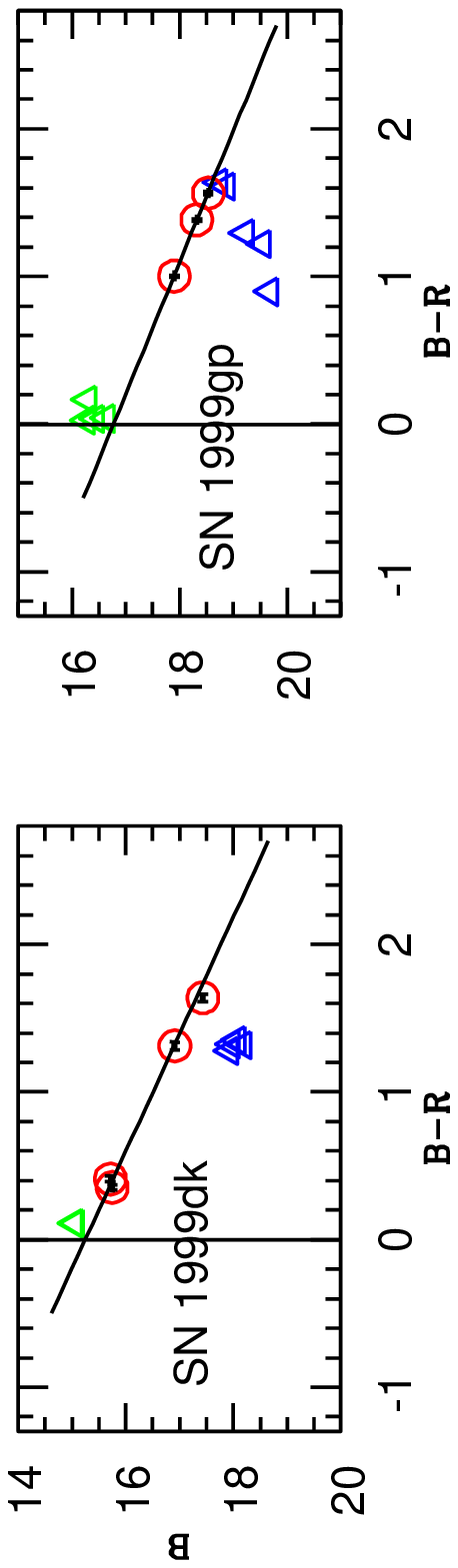}
\caption{Same as Fig. 1, but for B versus B-R.}
\end{figure}

\begin{figure}
\figurenum{A3}
\epsscale{1.0}
\plotone{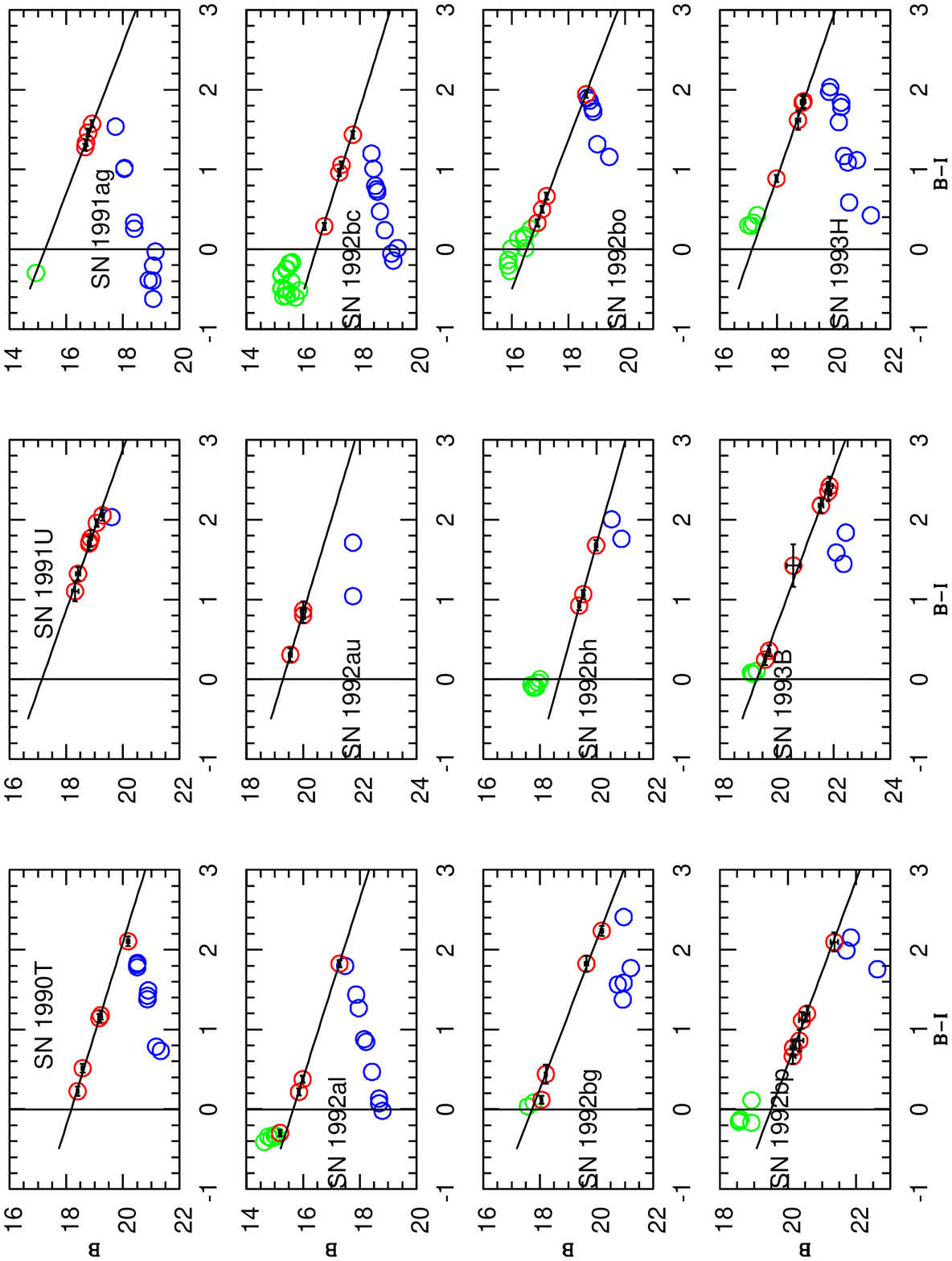}
\end{figure}

\begin{figure}
\figurenum{A3}
\epsscale{1.0}
\plotone{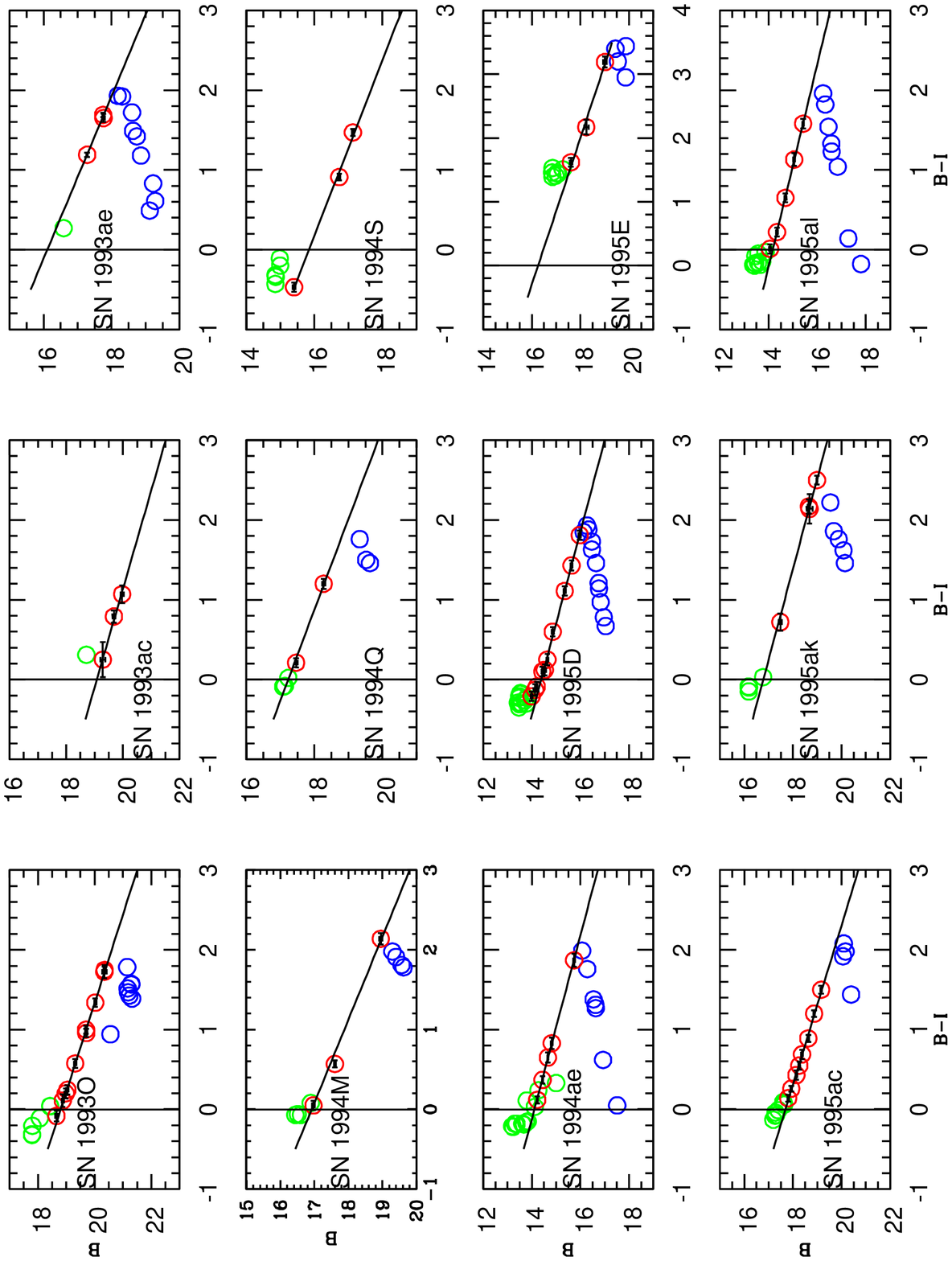}
\end{figure}

\begin{figure}
\figurenum{A3}
\epsscale{1.0}
\plotone{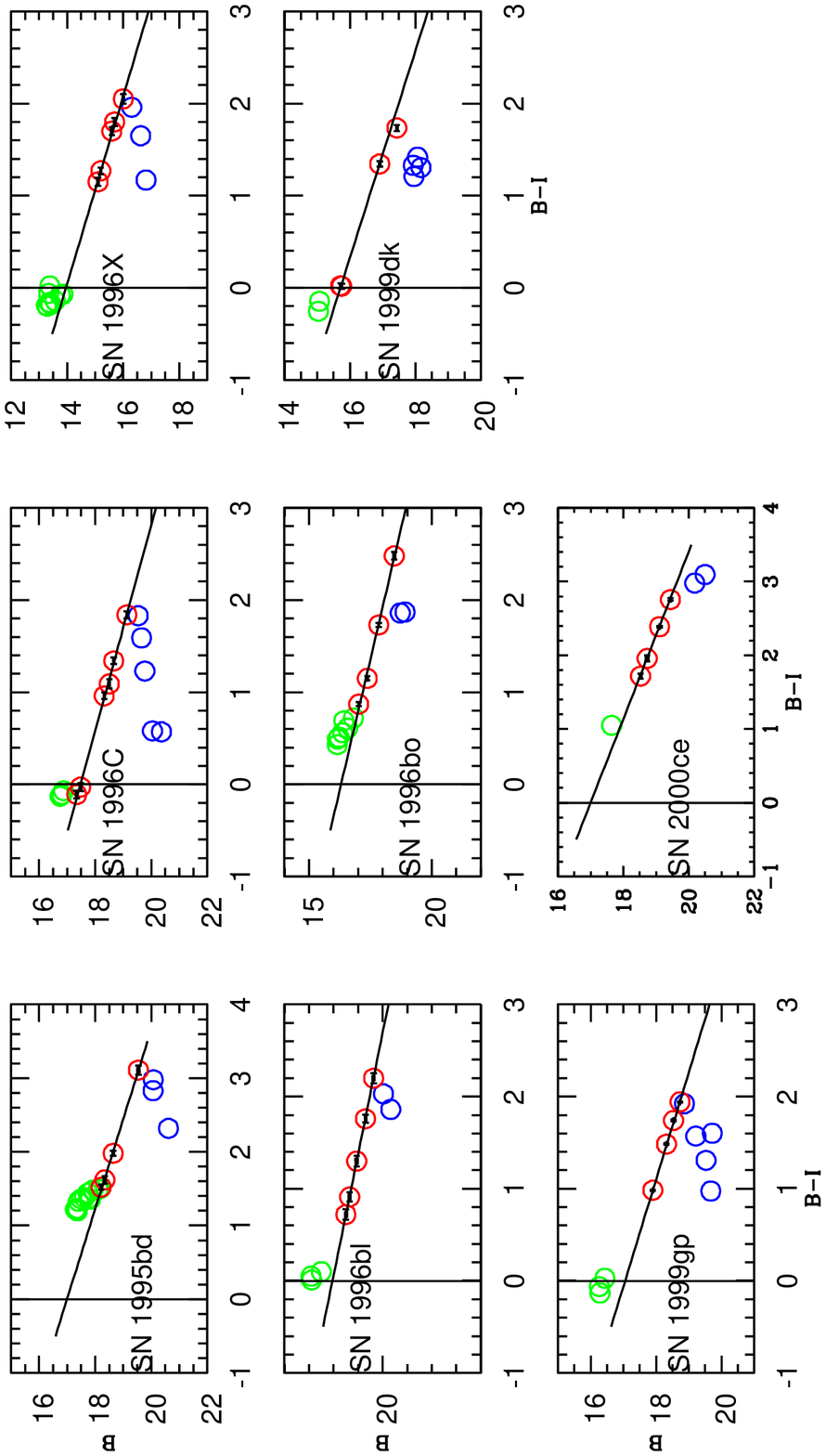}
\caption{Same as Fig. 1, but for B versus B-I.}
\end{figure}




\end{document}